\documentstyle[12pt,epsf,rotating]{article}

\catcode`@=11
\def\citer{\@ifnextchar [{\@tempswatrue\@citexr}{\@tempswafalse\@citexr[]}}

\def\@citexr[#1]#2{\if@filesw\immediate\write\@auxout{\string\citation{#2}}\fi
  \def\@citea{}\@cite{\@for\@citeb:=#2\do
    {\@citea\def\@citea{--\penalty\@m}\@ifundefined
       {b@\@citeb}{{\bf ?}\@warning
       {Citation `\@citeb' on page \thepage \space undefined}}%
\hbox{\csname b@\@citeb\endcsname}}}{#1}}
\catcode`@=12

\newcommand{\lsim}{\raisebox{-0.13cm}{~\shortstack{$<$ \\[-0.07cm] $\sim$}}~}

\newcommand{\ov}[1]{\overline{#1}}

\newcommand{\MeV}{\mbox{Me$\!$V}}
\newcommand{\GeV}{\mbox{Ge$\!$V}}
\newcommand{\TeV}{\mbox{Te$\!$V}}
\newcommand{\s}{\\ \vspace*{-3mm} }
\newcommand{\nn}{\noindent}
\newcommand{\non}{\nonumber}

\newcommand{\beq}{\begin{eqnarray*}}
\newcommand{\eeq}{\end{eqnarray*}}
\newcommand{\CP}{\mbox{${\cal CP}$}}
\newcommand{\SM}{\mbox{${\cal SM}$}}
\newcommand{\SUSY}{\mbox{${\cal SUSY}$}}
\newcommand{\MSSM}{\mbox{${\cal MSSM}$}}

\newcommand{\tg}{\mbox{tg}}
\newcommand{\tb}{\mbox{tg$\beta$}}

\newbox\mycount
\newcommand{\ctowidth}[2]{ \setbox\mycount=\hbox{$#2$}
                          \hbox to \wd\mycount{$ \hss #1 \hss $} }
\newcommand{\ltowidth}[2]{ \setbox\mycount=\hbox{$#2$}
                          \hbox to \wd\mycount{$\hskip0pt plus0pt minus1fil
                           #1 \hfill $} }
\newcommand{\rtowidth}[2]{ \setbox\mycount=\hbox{$#2$}
                          \hbox to \wd\mycount{$\hfill #1
                          \hskip0pt plus0pt minus1fil$} }

\begin{document}

\begin{titlepage}

\begin{flushright}
DESY 94--123\\
GPP--UdeM--TH--95--16\\
CERN--TH/95--30\\
hep-ph/9504378 \\
February 1995 \\
\end{flushright}

\vspace{1cm}

\begin{center}

{\large\sc {\bf HIGGS BOSON PRODUCTION AT THE LHC}}

\vspace{1cm}

{\sc M.~Spira$^1$, A.~Djouadi$^{2,3}$, D.~Graudenz$^4$, and
P.M.~Zerwas$^3$} \\
\vspace{1cm}

$^1$ II.Institut f\"ur Theoretische Physik, Universit\"at Hamburg, \\ Luruper
Chaussee 149, D--22761 Hamburg, FRG \\
\vspace{0.3cm}

$^2$ Groupe de Physique des Particules, Universit\'e de Montr\'{e}al, \\
Case 6128 A, H3C 3J7  Montr\'{e}al P.Q., Canada \\
\vspace{0.3cm}

$^3$ Deutsches Elektronen--Synchrotron DESY, D-22603 Hamburg, FRG \\
\vspace{0.3cm}

$^4$ Theoretical Physics Division, CERN, CH--1211 Geneva 23, Switzerland

\end{center}

\vspace{2cm}

\begin{abstract}
\normalsize
\noindent

 Gluon fusion is the main production mechanism for Higgs particles at
 the LHC. We present the QCD corrections to the fusion cross sections
 for the Higgs boson in the Standard Model, and for the neutral
 Higgs bosons in the minimal supersymmetric extension of the
 Standard Model. The QCD corrections are in general large and they
 increase the cross sections significantly. In two steps preceding
 the calculation of the production processes, we determine
 the QCD radiative corrections to Higgs decays into two photons
 and gluons.
 \end{abstract}

\end{titlepage}

\setcounter{page}{2}

\section[]{Introduction}
The Higgs mechanism is a cornerstone in the electroweak sector of the
Standard Model [$\cal SM$].
                        The fundamental particles, leptons,
quarks and gauge particles, acquire the masses through the interaction
with a scalar field \cite{P1}. To accomodate the well--established
electromagnetic and weak phenomena, this mechanism requires the
existence of at least one weak--isodoublet scalar field. After
absorbing three Goldstone modes to build up the longitudinal
$W^\pm_L, Z_L$ states, one degree of freedom is left over which
corresponds to a scalar particle. The properties of the Higgs
boson, decay widths and production mechanisms, can be
predicted if the mass of the particle is fixed \cite{P2}. \s

Even though the value of the Higgs mass cannot be predicted in the
Standard Model, constraints can nevertheless be derived from
internal consistency conditions \citer{P3,P4A}. Upper bounds on the
mass can be set by assuming that the Standard Model can be
extended up to a scale $\Lambda$ before perturbation breaks
down and new dynamical phenomena emerge. If the Higgs mass is less
than 200 \GeV, the Standard Model can be extended, with particles
interacting weakly, up to the GUT scale of order $10^{16}$ \GeV,
a prerequisite to the renormalization of $\sin^2 \theta_W$
from the symmetry value 3/8 down to $\sim$ 0.2 at low energies \cite{X1}.
For Higgs masses of more than about 700 \GeV, the theory
becomes strongly interacting already at energy scales in the
\TeV~region \cite{X2}. For the large top quark mass found
experimentally \citer{P5,P8A},
                the requirement of vacuum stability sets a lower
limit on the Higgs mass. For top masses of 150, 175 and 200 \GeV,
the lower limits on the Higgs mass are 40, 55 and 70 \GeV,
respectively, if the fields of the Standard Model become
strongly interacting at a scale of about 1 \TeV. The lower
limits are shifted upwards if the Standard Model with
weakly interacting fields extends up to energies of the
order of the Planck scale.
They decrease dramatically, however, if the vacuum is only assumed to
be metastable \cite{P4A}. \s

The most stringent experimental
                   limit on the Higgs mass in the $\SM$
has been set by LEP. A lower bound of 63.9 \GeV~has been found \cite{P7}
by exploiting the Bjorken process $Z \rightarrow Z^{\ast}H$
\cite{X3}. The
search will be extended to a Higgs mass near 80 to 90 \GeV~by
studying the Higgs--strahlung $e^+e^- \to Z^\ast \rightarrow ZH$ at LEP2
\cite{X4,X5}.
The detailed exploration of the Higgs sector in $e^+e^-$ collisions
for yet higher masses requires the construction of linear
colliders \cite{P8,habi}. \s

The search for Higgs particles after LEP2 will continue at
the $pp$ collider LHC \citer{P13A,P14A}. Several mechanisms contribute to the
production
of $\cal SM$ Higgs bosons in proton collisions \cite{habi}. The
dominant mechanism is the gluon fusion process \cite{P9}
\begin{displaymath}
pp \rightarrow  gg \rightarrow H
\end{displaymath}
which provides the largest production rate for the entire Higgs
mass range of interest. For large Higgs masses, the fusion
process $qq \to WW, ZZ \to H$ \cite{P10} becomes competitive, while
for Higgs particles in the intermediate mass range $M_Z  < M_H < 2M_Z$
the Higgs--strahlung off top quarks \cite{P11} and $W,Z$ gauge bosons
\cite{P12} are additional important production processes. \s

Rare decays to two photons will provide the main
                                            signature for the search
of $\cal SM$ Higgs particles in the lower part of the intermediate
range for masses below about 130 \GeV. To isolate the narrow
$\gamma \gamma$ signal in the huge $\gamma \gamma$ continuum
background, excellent energy and geometric resolution of the
$\gamma$ detectors is mandatory \cite{P14B,P14A}. Besides, excellent
$\mu$--vertex detectors may open the gate to the
dominant $b\overline{b}$ decay mode \cite{A1} even though the QCD jet
background remains very difficult to reject \cite{P14}. [At the expense
of considerably lower rates the background rejection can be
improved for both reactions by selecting Higgs--strahlung
events where additional isolated leptons from the
associated production of Higgs and top or $W$ bosons
reduce the QCD background.] Above this mass range,
Higgs decays to two $Z$ bosons -- one $Z$ being virtual
in the upper part of the intermediate range -- will be used to
tag the Higgs particle through $Z$ decays into pairs of charged
leptons \cite{P14B,P14A}. The background rejection
becomes increasingly simpler when the Higgs mass approaches
the real--$Z$ decay threshold. At the upper end of the
standard Higgs mass range of about 800 \GeV~the more frequent
decays of the $Z$ bosons into neutrino pairs and jets, as well as the
$WW$ decays of the
Higgs boson, with the $W$'s decaying to leptons and jets, must be
exploited to compensate
for the quickly dropping production cross section. \s
\vskip0.3cm

Supporting arguments for the supersymmetry extension
of the Standard Model are rooted in the Higgs sector.
Supersymmetric theories provide a natural mechanism for
retaining light
Higgs particles in the background of high GUT energy scales
\cite{P16}. In the minimal supersymmetric extension of the Standard Model
[$\cal MSSM$] two isodoublet scalar fields \cite{P24A} must be
introduced to preserve supersymmetry, leading to two
$\cal CP$--even neutral bosons $h^0$ and $H^0$,
a $\cal CP$--odd neutral boson $A^0$ and a pair of
charged Higgs bosons $H^\pm$. The observed value of
$\sin^2 \theta_W$ has been accurately predicted
in this theory \cite{P24B}, providing a strong motivation
for detailed studies of this theory \cite{A2}. \s

The mass of the lightest Higgs boson $h^0$ is bounded by the
$Z$ mass {\it modulo} radiative corrections of a few
tens of \GeV~\cite{P17,P26A}. [Triviality bounds similar to
the $\cal SM$ Higgs sector suggest an upper limit of
$\sim$ 150 \GeV~for supersymmetric theories in general
\cite{P18}.] The masses of the heavy neutral and
charged Higgs particles are expected to be in the range between
the electroweak symmetry breaking scale and the \TeV~scale. \s

Apart from radiative corrections the structure of the
$\cal MSSM$ Higgs sector is determined by two parameters,
one of the Higgs masses, in general $m_{A^0}$, and the
angle $\beta$ related to the ratio of the vacuum
expectation values of the two neutral Higgs fields, $\tg\beta
= v_2/v_1$. While
the overall strength of the couplings of the Higgs
bosons to the $\cal SM$ particles is given by the masses,
the mixing angles in the Higgs sector modify the
 hierarchy of the couplings
considerably. For example, the coupling of $h^0$ to
bottom quarks is strongly enhanced for large $\tg \beta$
compared with the coupling to the heavier top quarks.
Except for a small area in the [$m_{A^0}, \tg\beta$]
parameter space, $Z$ bosons couple predominantly
to $h^0$ while the complementary coupling to the heavy
$H^0$ Higgs boson is suppressed. The pseudoscalar
Higgs boson $A^0$ does not couple to the gauge
 bosons at the Born level. In addition, the Higgs particles
couple to the $\SUSY$ particles, with a strength,
however, which is essentially set by the gauge couplings. \s

The couplings determine the decay modes and therefore the
signatures of the Higgs particles. Apart from the small region
in the parameter space where the heavy Higgs boson $H^0$
decays into a pair of $Z$ bosons, rare $\gamma \gamma$ and
$\tau \tau$ decays must be utilized to search for the neutral
Higgs particles \cite{P14B,P14A} if $b$ quark decays cannot be
separated sufficiently well from the QCD background.
For large Higgs masses, decays into $\SUSY$ particles \cite{X6,X6A}
can provide additional experimental opportunities.
\s

The most important production mechanism for $\SUSY$ Higgs particles
at hadron colliders is the gluon fusion mechanism, similarly to the $\cal
SM$ Higgs boson production,
\begin{displaymath}
pp \rightarrow gg \rightarrow h^0, H^0, A^0
\end{displaymath}
and the Higgs radiation off top and bottom quarks. Higgs
radiation off $W/Z$ bosons and the $WW/ZZ$ fusion of
Higgs bosons play minor r\^oles in the
$\SUSY$ Higgs sector. \s
\vskip0.3cm

In the present analysis we have studied in detail the gluon
fusion of neutral
Higgs particles in the Standard Model and its
 minimal supersymmetric extension. The coupling of gluons to Higgs
bosons is mediated primarily
                   by heavy top quark loops, and eventually
bottom quark loops in supersymmetric theories. An extensive
literature already exists on various aspects of this mechanism. \s

The fusion mechanism has been proposed in Ref.~\cite{P9} for the production of
$\cal SM$ Higgs particles at hadron colliders, and has been discussed later in
great detail [see Ref.~\cite{P2,P13A,P14B,P14A} for a set of references].
The phenomenological issues for the production of Higgs particles in the
minimal supersymmetric extension of the Standard Model through the gluon fusion
mechanism were thoroughly discussed in Refs.~\cite{P19}.
All these analyses, however, were based on  lowest--order calculations. \s

Higher--order  QCD corrections have first been carried out in
Refs.~\cite{P20,P21} for the limit of large
loop--quark masses in the Standard Model. Later they were extended to the $\cal
MSSM$ Higgs spectrum \cite{P22,P23}; for this case, however, areas of the
parameter space in which $b$--quark loops are important, are not covered by the
approximation. The higher order QCD corrections of the fusion cross section for
the entire Higgs mass range have been given for the Standard Model in
Ref.~\cite{P24} and for its supersymmetric extension in Ref.~\cite{P25}. As
anticipated, the QCD corrections to the fusion processes are important and
experimentally significant. Quite generally they are positive and the
corresponding $K$ factors run up to values of $\sim 2$. \s

Besides the total production cross sections, the QCD corrected transverse
momentum spectra of the Higgs particles \cite{P26} as well as the cross
sections
for Higgs $+$ jet final states \cite{P27,P27A} are of great experimental
interest. \s

The theoretical analysis of QCD corrections to the gluon fusion of Higgs
particles involves complicated two--loop calculations; generic Feynman diagrams
are depicted in Fig.~1. Therefore they have first been performed for the
simpler
case of Higgs couplings to two photons, Fig.~2, for which the virtual QCD
corrections are a subset of the corrections to the Higgs couplings to gluons,
Fig.~3. In the experimentally relevant mass range, the QCD corrections to the
$\gamma \gamma$ widths of the $\SM$ and $\MSSM$  Higgs bosons are small, of
order $\alpha_s$ \cite{P22,P28}. In the $\MSSM$, special attention must be paid
to the kinematical range in which the heavy quark--antiquark threshold is
nearly
mass--degenerate with the pseudoscalar $A^0$ state so that non--perturbative
resonance effects must be controlled \cite{P29}. \s

The gluon decay width of the Standard Model Higgs particle has been
determined also in  next--to--leading order; the QCD corrections are positive
and numerically important \cite{P20,P30}. The QCD corrections to the rare
Higgs boson decay  $H \to Z\gamma$ [and to the reverse process $Z \to
H\gamma$] have been presented in Ref. \cite{MHZg}; in the mass ranges of
experimental interest they  are tiny, of order $\alpha_s$. [The
leading electroweak radiative corrections to the
$H gg$,  $H\gamma \gamma$ and $H \gamma Z$ couplings have been evaluated
in the heavy top quark limit to ${\cal O}(G_F m_t^2)$ \cite{X7}; they are
very small.] \s
\vskip0.3cm

This paper is divided into two parts. In the first part we will discuss the
gluon--gluon fusion cross section of the Higgs particle in the $\SM$ in
next--to--leading order QCD. The photonic and gluonic partial decay widths
of the particle are included in the first part of the discussion. The
calculations of the
production cross section and the decay widths have been performed for the
entire range of possible Higgs masses. The analytical results are summarized
in the
Appendix in terms of one--dimensional Feynman integrals. In the limit where
the Higgs mass is either small or large compared to the quark--loop masses,
the integration can be performed analytically and simple analytical results
can be derived for the production cross sections and the
decay widths. In the second part of the paper the analysis will be
extended to the ${\cal CP}$--even and ${\cal CP}$--odd neutral $\SUSY$ Higgs
bosons. To ensure a coherent presentation of the results, some material
published earlier in letter form will be included in the present
comprehensive report.

\vspace*{3mm}

\section{The Higgs Particle of the Standard Model}
\subsection{The Two--Photon Decay Width}
The two--photon decay width of the Higgs boson in the Standard Model,
\begin{displaymath}
H \rightarrow  \gamma \gamma
\end{displaymath}
is of interest for two reasons. In the lower part of the
intermediate mass range of the Higgs particle,
this rare decay mode provides the signature for the search
at hadron colliders \cite{P14B,P14A}. The
$\gamma \gamma$ width determines also the cross section for
Higgs production in $\gamma \gamma$ collisions \cite{P31}. Since
the $H \gamma \gamma$ coupling is mediated by triangle loops
of all charged particles, the precision measurement
of the $\gamma \gamma$ width eventually opens a window to particles
with masses much heavier than the Higgs mass.
If the masses of these particles are generated through the Higgs
mechanism, the couplings to the Higgs boson grow with the
masses, balancing the decrease of the triangle amplitude with
rising loop mass. As a result, the heavy particles do not
decouple. However, if the masses of the particles are
generated primarily by different mechanisms [as in supersymmetric
theories, for example], their effect on the $H \gamma \gamma$
coupling is in general small. \s

The decay process $H \rightarrow \gamma \gamma$ proceeds in the
Standard Model through $W$ and fermion loops, Fig.~2a,b.
Denoting the fermionic amplitude by $A_f$ and the $W$
contribution by $A_W$, the decay rate is determined by \cite{X4,P32}
\begin{equation}
\Gamma(H\rightarrow\gamma\gamma)=
               \frac{G_F\alpha^2m_H^3}{128\sqrt{2}\pi^3}
\left| \sum_f N_c Q_f^2 A_f(\tau_f) + A_W(\tau_W)
\right|^2
\end{equation}
where $N_c$ is the color factor, $Q_f$ the electric
charge of the fermion $f$. The scaling variables are defined by
\begin{eqnarray}
\tau_f = \frac{m^2_H}{4 m^2_f} \hspace{1cm}\mbox{and} \hspace{1cm}
\tau_W = \frac{m^2_H}{4 m^2_W}
\end{eqnarray}
The amplitudes $A_f$ and $A_W$ can be expressed as
\begin{eqnarray}
A_f(\tau) & = & 2 [\tau +(\tau -1)f(\tau)]/\tau^2  \nonumber \\
                                                            \nonumber \\
A_W(\tau) & = & - [2\tau^2 +3\tau+3(2\tau -1)f(\tau)]/\tau^2
\label{eq:6}
\end{eqnarray}
where the function $f(\tau)$ is given by
\begin{eqnarray}
f(\tau)=\left\{
\begin{array}{ll}  \displaystyle
\arcsin^2\sqrt{\tau} & \tau\leq 1 \\
\displaystyle -\frac{1}{4}\left[ \log\frac{1+\sqrt{1-\tau^{-1}}}
{1-\sqrt{1-\tau^{-1}}}-i\pi \right]^2 \hspace{0.5cm} & \tau>1
\end{array} \right.
\label{eq:ftau}
\end{eqnarray}

If the Higgs mass is smaller than the $WW$ and
$f \bar f$ pair thresholds, the amplitudes are
real; above the thresholds they are complex,
Fig.~4. Below the
      thresholds the $W$ amplitude is always
dominant, falling from $(-7)$ for very light Higgs masses
to $(-5 - 3 \pi^2/4)$ at the $WW$ threshold; for large Higgs masses
the $W$ amplitude approaches $A_W \to (-2)$.
Quark contributions increase from
4/3 for light Higgs masses (compared with the quark mass)
to 2 at the quark--antiquark threshold; far above the fermion
threshold, the amplitude vanishes linearly in $\tau$ {\it mod.}
logarithmic coefficients,
$A_f \to -[\log(4\tau)-i\pi]^2/2\tau$, i.e.~proportional to $m_f^2/m_H^2$.
The contribution
of the $W$ loop interferes destructively with the
quark loop. For Higgs masses of about 600 \GeV, the two
contributions nearly cancel each other \cite{P33}.\s

Since the $Hff$ coupling is proportional to the fermion mass, the contribution
of light fermions is negligible so that in the Standard Model with three
families, only the top quark and the $W$ gauge boson effectively contribute to
the $\gamma \gamma$ width.  Since the $W$ and fermion loops interfere
destructively, a fourth generation of heavy fermions would reduce the size of
the $H \gamma \gamma$ coupling.  For small Higgs masses the additional
contributions of the heavy quarks and the charged lepton would suppress
the decay width by about one order of magnitude.  \s

To fully exploit the potential of the $\gamma \gamma$ decay
mode of the Higgs particle and the production
in $\gamma \gamma$ collisions, the QCD corrections must
be shown to be under proper control. To include the gluonic
QCD corrections, twelve two--loop diagrams plus the
associated counter terms must be taken into account.
Generic examples are depicted in Fig.~2c. \s

Throughout this analysis we have adopted the on--shell renormalization
scheme which is convenient for heavy quarks. In this scheme
the quark mass $m_Q$ is defined as the pole of the
propagator\footnote{We have chosen $m_t = 174~\GeV$ for
the $t$ pole mass \cite{P6} and $m_b = 5~\GeV$ for the $b$ pole mass.},
related in the following way to the running mass
\begin{equation}
m_Q(\mu_Q^2) = m_Q \left[ \frac{\alpha_s(\mu_Q^2)}{\alpha_s(m_Q^2)}
\right]^{12/(33-2N_F)} \{ 1+{\cal O} (\alpha_s^2) \}
\label{eq:mrun}
\end{equation}
at the mass renormalization point $\mu_Q$. It should be noted that this
definition of the running mass does {\it not} coincide with the running
$\overline{MS}$ mass.  The wave function is renormalized
[$Z_2^{1/2}$] such that the residue at the pole is equal to unity.  The
photon--quark vertex is renormalized at zero--momentum transfer;  the standard
QED Ward identity renders the corresponding renormalization factor equal to
the renormalization factor of the wave function.  Since the fermion masses are
generated in the Standard Model by the interaction with the Higgs field, the
renormalization factor associated with the Higgs--quark vertex [$Z_{HQQ}$] is
fixed unambiguously by the renormalization factors $Z_m$ for the mass and
$Z_2$ for the wave function.  From the Lagrangian
\begin{eqnarray}
{\cal L}_0 & = & -m_0 \bar Q_0 Q_0 \frac{H}{v} \non \\
           & = & -m_Q \bar Q Q \frac{H}{v}
                 + Z_{HQQ} m_Q \bar Q Q \frac{H}{v}
\end{eqnarray}
we find \cite{P35}
\begin{equation}
Z_{HQQ} = 1-Z_2 Z_m
\end{equation}
In contrast to the renormalized photon--fermion vertex, the scalar $HQQ$ vertex
$\Gamma (p', p)$ is renormalized at zero momentum transfer by a finite amount
$\gamma_m$ of order $\alpha_s$ after subtracting $Z_{HQQ}$ due to the lack of
a corresponding Ward identity.  The finite renormalization $\gamma_m$
corresponds to the anomalous mass dimension discussed later.  \s

We have calculated the two--loop amplitudes using dimensional regularization.
The five--dimensional Feynman parameter integrals of the amplitudes have been
reduced analytically down to one--dimensional integrals over polylogarithms
\cite{P37} which have been evaluated numerically\footnote{The scalar integral
associated with the gluon correction to the $HQQ$ vertex has also been
analyzed by means of analytical \cite{Mmainz} and novel approximation methods
\cite{Mbielefeld}.  The results are in agreement within an accuracy of
$10^{-5}$.} [see Appendix A].  In the two limits where $m^2_H/4m^2_Q$ is
either very small or very large, the amplitudes could be calculated
analytically.  \s

The QCD corrections of the quark contribution to the
two--photon Higgs decay amplitude can be parameterized as
\begin{equation}
A_Q = A_Q^{LO} \left[ 1+ C_H \frac{\alpha_s}{\pi} \right]
\end{equation}
The coefficient $C_H$ splits into two parts,
\begin{equation}
C_H = c_1 + c_2 \log \frac{\mu_Q^2}{m_Q^2}
\label{eq:c1c2}
\end{equation}
where the functions $c_i$ depend only on the scaling variable
$\tau = m^2_H/4m^2_Q
(\mu_Q^2)$. The {\it same} running quark mass $m_Q (\mu_Q^2)$,
evaluated at the renormalization scale $\mu_Q$, enters
in the lowest--order triangle amplitude $A^{LO}_Q$.
The scale in $\alpha_s$ is arbitrary to this order; however, in
practice it should be defined of order $m_H$. As a typical
renormalization scale we have chosen $\mu_Q = m_H/2$.
This choice suggests itself for two reasons. The $Q \overline{Q}$
decay threshold is [perturbatively] defined at the correct
position $2m_Q(m_Q) = 2m_Q$. In addition, it turns out
{\it a posteriori} that all relevant large logarithms are effectively
absorbed into the running mass for the entire
physically interesting range of the scaling variable
$\tau$. \s

The correction factor $C_H$ is displayed in Fig.~5, illustrating the preferred
choice $\mu_Q = m_H/2$ for the renormalization scale.  The coefficient is real
below the quark threshold and complex above.  Near the threshold, within a
margin of a few \GeV, the present perturbative analysis is not valid.  The
formation of a $P$--wave $0^{++}$ resonance, interrupted however by the rapid
quark decay \cite{P38}, modifies the amplitude in this range \cite{P29}.  The
perturbative analysis may nevertheless account for the resonance effects in a
dual way.  Since $Q \overline{Q}$ pairs cannot form $0^{++}$ states {\it at}
the threshold, $\Im m \, C_H$ vanishes there.  $\Re e C_H$ develops a maximum
very close to the threshold.  \s

The QCD--corrected $\gamma \gamma$ decay width of the
Higgs boson is shown in Fig.~6a. The correction
relative to the lowest order is small in general,
Fig.~6b. The corrections are seemingly large
only in the area where the destructive $W$-- and $Q$--loop
interference makes the decay amplitude nearly vanish. \s

\vspace*{2mm}

\noindent
\underline{\it The Limit of Large Quark--Loop Mass}

\vspace*{6mm}

\nn
In the limit $m^2_H/4m^2_Q \rightarrow 0$, the five--dimensional Feynman
parameter integrals can be evaluated analytically.  The correction factor for
the $H \gamma \gamma$ coupling
\begin{eqnarray}
m_H^2/4m_Q^2 \to 0\,: \hspace{0.5cm} 1+
C_H \frac{\alpha_s}{\pi} \to  1 - \frac{\alpha_s}{\pi}
\end{eqnarray}
agrees with the result of the numerical integration in this limit.  \s

The $H \gamma \gamma$ coupling can also be derived by means of
a general low--energy theorem for amplitudes involving soft
Higgs particles \cite{X4,P32},
$\lim_{p_H\to 0} {\cal A} (XH) = (m_0 / v) \partial {\cal A} (X) /
\partial m_0$. The theorem is easy to prove.
For zero 4--momentum the kinetic
 derivative term in the Lagrangian can be neglected
 and the [space--time independent] Higgs field can
 be incorporated by adding the potential energy to the bare mass term,
$ m_0 \to m_0 ( 1+H/v)$,
in the Lagrangian. The expansion of the bare propagators for small
values of $H/v$ is then equivalent to inserting a
zero--momentum Higgs field in an arbitrary amplitude
${\cal A} (X)$,
\begin{eqnarray}
\frac{1}{\not  \! k - m_0} \rightarrow
\frac{1}{\not  \! k - m_0} \frac{m_0}{v}  \, \, \frac{1}
{\not  \! k - m_0}
\end{eqnarray}
and generating this way the amplitude ${\cal A} (XH)$.  Since the bare mass
$m_0$ and the renormalized mass $m_Q$ are related by the anomalous mass
dimension, $ d \log m_0 = (1 + \gamma_m) d \log m_Q$, we find for the final
form of the theorem
\begin{equation}
\lim_{p_H \to 0} {\cal A} (XH) = \frac{1}{1+\gamma_m} \frac{m_Q}{v}
\frac{\partial}{\partial m_Q} {\cal A} (X)
\end{equation}

It is well--known that the theorem can be exploited to derive the $H \gamma
\gamma$ coupling in lowest order \cite{X4,P32}.  However, the theorem is also
valid if radiative QCD corrections are taken into account.  For large fermion
masses, the vacuum polarization of the photon propagator at zero momentum is
given by
\begin{eqnarray}
\Pi = - e_Q^2\frac{\alpha}{\pi} \Gamma(\epsilon)
\left(\frac{4\pi\mu^2}{m_Q^2}\right)^\epsilon \left[ 1+\frac{\alpha_s}{2\pi}
\Gamma(1+\epsilon) \left(\frac{4\pi\mu^2}{m_Q^2}\right)^\epsilon
+ {\cal O} (\epsilon) \right]
\end{eqnarray}
so that
\begin{eqnarray}
 m_Q \frac{\partial \Pi}{\partial m_Q} = 2 \frac{\alpha}{\pi}
\left( 1+\frac{\alpha_s}{\pi} \right)
\end{eqnarray}
{}From the anomalous mass dimension to lowest order,
\begin{eqnarray}
\gamma_m = 2\alpha_s/\pi
\end{eqnarray}
one readily derives the correction $C_H$ of the
$H \gamma \gamma$ coupling
\begin{eqnarray}
m_H^2/4m_Q^2 \to 0\,: \hspace{0.5cm} 1+
C_H\frac{\alpha_s}{\pi} \to \frac{1+\alpha_s/\pi} {1+2 \alpha_s/\pi}
\ = 1 - \frac{\alpha_s}{\pi}
\end{eqnarray}
Compared with the radiative QCD correction to the photon propagator,
$(1 + \alpha_s / \pi)$, just the sign of the
correction is reversed, $(1- \alpha_s/\pi)$, for the
$H\gamma \gamma$ coupling \cite{P28}. In the notation of eq.(\ref{eq:c1c2})
the correction is attributed to $c_1$ while $c_2
\sim 1/m^2_Q$ vanishes for large quark masses. \s

The same result can be derived by exploiting well--known results on
the anomaly in the trace of the energy--momentum tensor
\cite{P36},
\begin{equation}
\Theta_{\mu\mu} = (1+\gamma_m) m_0 \overline{Q}_0 Q_0 +
\frac{1}{4}\frac{\beta_\alpha}{\alpha} F_{\mu\nu} F_{\mu\nu}
\end{equation}
$\beta_\alpha$ denotes the mixed QED/QCD
$\beta$ function defined by $\partial
\alpha(\mu^2)/\partial \log \mu = \beta_\alpha$.
Since the matrix element $\langle \gamma \gamma | \Theta_{\mu \mu} |0 \rangle$
vanishes for infrared photons, the coupling of the two--photon
state to the Higgs source $(m_0/v) \overline{Q}_0 Q_0$
is given by $\beta'_\alpha/[4\alpha (1 + \gamma_m)]$,
with $\beta'_\alpha = 2 e_Q^2 \alpha^2/\pi (1 + \alpha_s/\pi)$
including {\it only} the heavy quark contribution to the QED/QCD $\beta$
function.  Thus the $H \gamma \gamma$ coupling is described by the
effective Lagrangian
\begin{equation}
{\cal L} (H\gamma\gamma) = \frac{e_Q^2\alpha}{2\pi}~\left(
\sqrt{2}G_F\right)^{1/2}~\left[ 1-\frac{\alpha_s}{\pi} \right] F_{\mu\nu}
F_{\mu\nu}~H
\label{eq:LHgam}
\end{equation}
which is apparently equivalent to the previous derivation of the $H \gamma
\gamma$ coupling in the limit $m^2_H/4m^2_Q \rightarrow 0$.  \s

\vspace*{2mm}

\noindent
\underline{\it The Limit of Small Quark--Loop Masses} \s

\vspace*{2mm}

\nn
In the limit $m_Q (\mu_Q^2) \rightarrow 0$ the leading and subleading
logarithms of the
QCD correction $C_H$ can be evaluated analytically:
\begin{equation}
m_Q(\mu_Q^2) \to 0\,: \hspace{0.5cm} C_H \to -\frac{1}{18}
\log^2 (-4\tau - i\epsilon) - \frac{2}{3}
\log (-4\tau -i\epsilon) + 2\log\frac{\mu_Q^2}{m_Q^2}
\end{equation}
and, split into real and imaginary parts,
\begin{eqnarray}
\Im m C_H & \to & \frac{\pi}{3} \left[ \frac{1}{3} \log(4\tau) + 2
\right] \nonumber \\
 \Re e C_H & \to & -\frac{1}{18}\left[\log^2 (4\tau)-\pi^2 \right]
-\frac{2}{3}\log(4\tau) +2 \log \frac{\mu_Q^2}{m_Q^2}
\label{eq:1.1.14}
\end{eqnarray}
The choice of the renormalization scale $\mu_Q$ is crucial for the
size of $C_H$. Choosing the on--shell definition
$\mu_Q = m_Q$ leads to very large corrections in the imaginary
as well as the real part, as demonstrated in
Fig.~5. By contrast, for $\mu_Q = \frac{1}{2} m_H$,
the corrections in the real and imaginary part remain small in
the entire $\tau$ range of interest, $\tau \lsim \mbox{a few~} \times
10^4$ for $m_b \sim $ 3 \GeV~and $m_H \lsim 1~\TeV$.
[This coincides with the corresponding
observation for the decay $H \rightarrow b \overline{b}$
where the running of the $b$--mass up to the scale
$\frac{1}{2} m_H$ absorbs the leading logarithmic
coefficients \cite{P35}.] Only for
log $\tau$ values above the physical range must the leading logarithmic
corrections be summed up;
such an analysis is beyond the scope of the present investigation.

\subsection{The Gluonic Decay Width}
Gluonic decays of the Higgs boson
\begin{displaymath}
H \to gg
\end{displaymath}
are of physical interest for arguments similar to the preceding
section. However, there are some qualitative differences. Since the
particle loops mediating the $Hgg$ coupling carry color charges,
the color--neutral $W,Z$ gauge bosons do not
contribute. The gluonic branching ratio can only be measured
directly at $e^+e^-$ colliders and for Higgs masses
presumably less than about 140 \GeV~\cite{P8} since it drops
quickly to a level below $10^{-3}$ for increasing masses.
In this range, a fourth generation of fermions would
enhance the branching ratio to a level where it becomes
competitive with the dominant $b\bar{b}$ decay mode. \s

The gluonic width determines the production cross section
of Higgs bosons in gluon--gluon fusion to leading order at hadron
colliders. The cross section, however, is strongly
affected by QCD radiative corrections so that the
width can be measured in this indirect way only within
about 20\%. \s

At the Born level the contribution of heavy quarks to
the gluonic width in the Standard Model is given by
\begin{eqnarray}
\Gamma(H\rightarrow gg) = \frac{G_F \alpha_s^2}{36\sqrt{2}\pi^3} m_H^3
\left| \frac{3}{4} \sum_Q A_Q(\tau_Q) \right|^2
\end{eqnarray}
where $A_Q$ denotes the quark amplitude, already discussed in
eq.(\ref{eq:6}), without the color factor.
The top quark contribution is by far
dominant in the $\SM$. Any additional
heavy quark from a fourth family etc. increases the
decay amplitude by a factor 2 in the limit where
the Higgs mass is small compared with the $Q \overline{Q}$
threshold energy. \s

The QCD corrections to the gluonic decay width \cite{P20,P30} are
large. Several classes of diagrams must be calculated
in addition to those familiar from the two--photon
decay amplitude. Generic examples are shown in
Fig.~3. The virtual corrections involve the non--abelian three--gluon
and four--gluon couplings, and the counter terms associated with the
renormalization $Z_g - 1 = (Z_1 - 1) - \frac{3}{2} (Z_3 - 1)$ of
the QCD coupling. We have defined $\alpha_s$ in the
                                                 $\overline{MS}$
scheme with five active quark flavors and the heavy top quark
decoupled \cite{Malpdec}.
Besides the virtual corrections, three--gluon and gluon
plus quark--antiquark final states must be taken into
account,
\begin{displaymath}
H\to ggg \hspace{0.5cm} \mbox{and} \hspace{0.5cm} gq\bar q
\label{eq:1.2.2}
\end{displaymath}
In the quark channel we will restrict ourselves to
the light quark species which we will
treat as massless particles\footnote{We include $c,b$ quarks among the
light quarks so that all large logarithms $\log m_H^2/m_{c,b}^2$,
associated with final state particle splitting,
are removed by virtue of the Kinoshita--Lee--Nauenberg theorem.
This assumes that when the theoretical prediction will be compared
with data,  $c$ and $b$ quark final states in collinear
configurations are not subtracted. Note that in Higgs decays
to $c,b$ quark pairs plus an additional gluon jet,
the heavy quarks are emitted preferentially back--to--back and
not in collinear configurations. [A more detailed phenomenological
analysis of these final states is in progress.]}.
As a consequence of chirality
conservation, the gluon decay amplitude does not
interfere in this limit with the amplitude in which the
$q \overline{q}$ pair is coupled directly to the
Higgs boson [$g (Hqq)$ of order $m_q$, but kept non--zero].
This would be different for top quark decays
$H \rightarrow t \overline{t}g$ where the decay
mechanism of Fig.~3, however, is a higher--order
effect, suppressed to ${\cal O} (g_s^2)$ already
at the amplitude level with respect to the gluon bremsstrahlung
correction of the basic $t\bar t$ decay amplitude.
The light--quark final states in the QCD corrections to the
 gluonic decays, on the other hand, must be taken into
account since they are energy--degenerate with the
gluon final states. \s

The result can be written in the form
\begin{equation}
\Gamma(H\rightarrow gg(g),~gq\bar q) = \Gamma_{LO}(H\rightarrow gg)
\left[ 1 + E(\tau_Q) \frac{\alpha_s}{\pi} \right]
\end{equation}
with
\begin{equation}
E(\tau) = \frac{95}{4} - \frac{7}{6} N_F
+ \frac{33-2N_F}{6}\ \log \frac{\mu^2}{m_H^2} + \Delta E
\end{equation}
The first three terms survive in the limit of large loop masses while $\Delta
E$ vanishes in this limit.  $\mu$ is the renormalization point and defines the
scale parameter of $\alpha_s$.  It turns out {\it a posteriori} that the
higher order corrections are minimized by choosing the pole mass $m_Q$ for the
renormalized quark mass;  this is evident from Fig.~7a.  The correction
$\Delta E$, given explicitely in the Appendix, is displayed in Fig.~7b for the
physically relevant mass range.  In Fig.~8 we present the gluonic width of the
Higgs boson including the QCD radiative corrections\footnote{In all numerical
analyses and figures, the contributions of the $b$ quark loops have been
included.  Even for small Higgs masses these effects remain less than about
10\% of the leading $t$ quark contributions.}.  \s

The total decay width and the branching ratios of all decay processes
in the Standard Model are shown in Fig.~9 for Higgs boson masses up to 1 \TeV.
All known QCD and leading electroweak radiative corrections are included. \s

The size of the QCD radiative corrections depends on the choice
of the renormalization scale $\mu$ for any fixed order
of the perturbative expansion. A transparent prescription
is provided by the BLM scheme \cite{P40} in which the
$N_F$ dependent coefficient of the correction
is mapped into the coupling
$\alpha_s$, summing up quark and gluon loops in the
gluon propagators. We shall apply this prescription in
the large loop--mass limit where the amplitude can be calculated
analytically:
\begin{equation}
m_H^2/4m_Q^2 \to 0: \hspace{1cm} E(\tau) = \frac{95}{4} - \frac{7}{6} N_F
+ \frac{33-2N_F}{6}\ \log \frac{\mu^2}{m_H^2}
\label{eq:SMEinf}
\end{equation}
Choosing
\begin{equation}
\mu_{BLM} = e^{-\frac{7}{4}} m_H \approx 0.17 m_H
\end{equation}
the $N_F$ dependent part drops out of $E(\tau)$ and we are left with
\begin{equation}
\Gamma(H\rightarrow gg(g) + gq\bar q) = \Gamma_{B}[\alpha_s(\mu_{BLM})]
\left[ 1 + \frac{9}{2} \frac{\alpha_s}{\pi} \right]
\label{eq:1.2.6}
\end{equation}
A large fraction of the total QCD correction is thus
to be attributed to the
renormalization of the coupling. \s

We shall conclude this subsection with a few comments on the
effective $Hgg$ Lagrangian \cite{P21}. In the same way as for
$H \gamma \gamma$, we can derive the effective gluon
Lagrangian for quark--loop masses large compared to the
Higgs mass by taking the derivative of the gluon
propagator with respect to the bare quark mass for
$q^2 = 0$. Introducing again the anomalous mass
dimension $\gamma_m$, one finds for the gauge--invariant
Lagrangian,
\begin{equation}
{\cal L}_{Hgg} = \frac{1}{4}\ \frac{\beta (\alpha_s) }
{1+\gamma_m(\alpha_s)}\
G^a_{\mu\nu} G^{a}_{\mu\nu} \frac{H}{v}
\end{equation}
where

\begin{equation}
\beta = \frac{\alpha_s}{3\pi}\left[
1+\frac{19}{4}\frac{\alpha_s}{\pi}\right] \hspace{1cm} \mbox{and}
\hspace{1cm} \gamma_m = 2 \frac{\alpha_s}{\pi}
\end{equation}

\nn so that to second order
\begin{equation}
{\cal L}_{Hgg} = \frac{\alpha_s}{12\pi}~\left( \sqrt{2} G_F \right)^{1/2}
\left[ 1+\frac{11}{4}\ \frac{\alpha_s}{\pi} \right] G^a_{\mu\nu}
G^{a}_{\mu\nu}~H
\label{eq:LHgg}
\end{equation}

\nn As a consequence of the non--abelian gauge invariance the
Lagrangian describes, besides the $Hgg$ coupling, also
the $Hggg$ and $Hgggg$ couplings, Fig.~10a. \s

In contrast to the effective $H \gamma \gamma$
Lagrangian, ${\cal L}_{Hgg}$ does not describe the
$Hgg$ interaction to second order in $\alpha_s$ in total.
This Lagrangian accounts only for the interactions mediated by the heavy
quarks directly, but it does not include the quantum effects of the
light fields\footnote{Technically, the additional contributions are
proportional to a common factor $(\mu^2/m_H^2)^\epsilon$
which vanishes if the Higgs mass is set to zero before $\epsilon$ is
driven to $(-0)$. [Note that these mass singularities
are regularized formally for $\epsilon < 0$.] However,
keeping the Higgs mass non--zero but small, the expansion in
$\epsilon$ gives rise to $\log \mu/m_H$ terms which fix the
size of the renormalization scale of the physical process.}:
${\cal L}_{Hgg}$ must be added to the light--quark and gluon part of
the basic QCD Lagrangian, and this sum then serves as a new effective
 Lagrangian for Higgs--gluon--light quark interactions. Physical
observables associated with the low--energy Higgs particle are
calculated by means of this effective Lagrangian in the standard
way, generating gluon self--energies, vertex corrections,
gluon--by--gluon scattering, gluon splitting to gluon and light
quark pairs, etc. {\it In summa}, the diagrams displayed in Fig.~10b
must be evaluated, taking into account also the corresponding counter
terms that renormalize the coupling $\alpha_s$ and the
gluon wave function. \s

The fixed--order program discussed so far  can be applied to the mass
region where $m_H/2m_Q$ is small [in essence $< 1$] \cite{P20,P21,P41}
but $\log m_Q/m_H$ still moderate so that logarithmic terms
 need not be summed up. This is the kinematical region of physical
interest. Based on a careful RG analysis \cite{P30}, the
logarithmic terms have been summed up in the limit where also
 $\log m_Q/m_H$ is large. This leads to the
plausible result that the energy scale in the effective
$Hgg$ Lagrangian is set by the heavy--quark mass while the
Higgs mass is the scale relevant for the additional light--quantum
 fluctuations. This can be incorporated by substituting
$\alpha_s E(\tau) \to [11/2] \alpha_s(m_Q) + [73/4-7/6 N_F]
\alpha_s(m_H)$ in eq.(\ref{eq:SMEinf}), leaving us with the light--quantum
 fluctuations as the main component of the QCD corrections in this
mathematical limit. For moderate values of $\log m_Q/m_H$ the
splitting is of higher order in the QCD coupling and can be
neglected.

\bigskip

\subsection{Higgs Boson Production in $pp$ Collisions}

Gluon fusion \cite{P9}
\begin{displaymath}
pp \to gg \to H
\end{displaymath}
\nn
is the main production mechanism of Higgs bosons in high--energy
$pp$ collisions throughout the entire
Higgs mass range. As
discussed before, the gluon coupling to the Higgs boson in the
Standard Model is mediated by triangular loops of top quarks.
The decreasing form factor with rising loop mass is
counterbalanced by the linear growth of the Higgs coupling
with the quark mass. [Heavier quarks still, in a fourth
family for instance, would add the same contribution to
the production amplitude if their masses were generated
through the standard Higgs mechanism.] \s

To lowest order the parton cross section, Fig.~1a, can
be expressed by the gluonic width of the Higgs boson,
\begin{eqnarray}
\hat\sigma_{LO} (gg\to H) & = & \frac{\sigma_0}{m_H^2} \delta
(\hat s -m_H^2) \\
\sigma_0 & = & \frac{8\pi^2}{ m_H^3} \Gamma_{LO} (H\to gg) \non
\end{eqnarray}
where $\hat{s}$ is the $gg$ invariant energy squared.
Recalling the lowest--order two--gluon decay width of the Higgs
boson, we find
\begin{equation}
\sigma_0 = \frac{G_{F}\alpha_{s}^{2}(\mu^2)}{288 \sqrt{2}\pi} \
\left| \ \frac{3}{4} \sum_{q} A_Q (\tau_{Q}) \ \right|^{2}
\end{equation}
The $\tau_Q$ dependence of the form factor has been given in
eq.(\ref{eq:6}). With rising mass, the width of the $\SM$ Higgs boson
quickly becomes broader. This effect can be incorporated in
the lowest--order approximation by substituting the Breit--Wigner
form for the zero--width $\delta$--distribution
\begin{eqnarray}
\delta(\hat s - m_H^2) \to \frac{1}{\pi}~\frac{\hat s \Gamma_H/m_H}
{(\hat s - m_H^2)^2 + (\hat s \Gamma_H/m_H)^2}
\end{eqnarray}
and changing kinematical factors $m^2_H \rightarrow \hat{s}$
appropriately. \s

Denoting the gluon luminosity as
\begin{equation}
\frac{d{\cal L}^{gg}}{d\tau} = \int_\tau^1 \frac{dx}{x}~g(x,M^2)
g(\tau /x,M^2)
\end{equation}
\nn the lowest--order proton--proton cross section is
found in the narrow--width approximation to be
\begin{equation}
\sigma_{LO}(pp\to H) = \sigma_0 \tau_H \frac{d{\cal L}^{gg}}{d\tau_H}
\end{equation}
\nn where the Drell--Yan variable is defined, as usual, by
\begin{eqnarray}
\tau_H = \frac{m^2_H}{s}
\end{eqnarray}
with $s$ being the invariant $pp$ collider energy squared. The
expression $\tau_H d {\cal L}^{gg}/d \tau_H$ is only
mildly divergent for $\tau_H \rightarrow 0$. \s

The QCD corrections to the fusion process $gg \rightarrow H$
\cite{P20,P21,P24}, Fig.~1b,
\begin{displaymath}
gg \rightarrow H(g) \hspace{0.5cm} \mbox{and} \hspace{0.5cm}
gq \rightarrow Hq,~q\overline{q} \rightarrow Hg
\end{displaymath}
involve the virtual corrections for the $gg \to H$ subprocess and the
radiation of gluons in the final state; in addition, Higgs
bosons can be produced in gluon--quark collisions and
quark--antiquark annihilation. These subprocesses
contribute to the Higgs production at the same order of
$\alpha_s$. The virtual corrections modify the lowest--order
fusion cross section by a coefficient linear in $\alpha_s$.
Gluon radiation leads to 2--parton final states with invariant
energy $\hat{s} \ge m_H^2$ in the $gg, gq$ and $q\overline{q}$
channels. The parton cross sections for the subprocess $i + j
\rightarrow H + X$ may thus be written
\begin{equation}
\hat\sigma_{ij} = \sigma_0 \left\{
\delta_{ig}\delta_{jg}\left[ 1+C(\tau_Q)\frac{\alpha_s}{\pi} \right]
\delta(1-\hat{\tau}) + D_{ij}(\hat{\tau},\tau_Q) \frac{\alpha_s}{\pi}
\Theta (1- \hat{\tau}) \right\}
\label{eq:sigij}
\end{equation}
for $i, j = g, q, \overline{q}$. The new scaling variable
$\hat{\tau}$, supplementing the variables $\tau_H=m_H^2/s$ and
 $\tau_Q=m_H^2/4m_Q^2$ introduced earlier, is defined at the parton level,
\begin{eqnarray}
\hat{\tau} = \frac{m^2_H}{\hat{s}}
\end{eqnarray}
The quark--loop mass is defined as the pole mass in the scaling variable
$\tau_Q$.  The coefficients $C(\tau_Q)$ and $D_{ij} (\hat{\tau},\tau_Q)$ have
been determined by means of the same techniques as described for the $H \gamma
\gamma$ and $Hgg$ couplings at great detail.  The lengthy analytic expressions
for arbitrary Higgs boson and quark masses are given in the Appendix in the
form of one--dimensional Feynman integrals.  The quark--loop mass has been
defined in the on--shell renormalization scheme, while the QCD coupling is
taken in the $ \overline{MS}$ scheme.  If all the corrections (\ref{eq:sigij})
are added up, ultraviolet and infrared divergences cancel.  However collinear
singularities are left over.  These singularities are absorbed into the
renormalization of the parton densities \cite{P42}.  We have adopted the
$\overline{MS}$ factorization scheme for the renormalization of the parton
densities.  The final result for the $pp$ cross section can be cast into the
form
\begin{equation}
\sigma(pp \rightarrow H+X) = \sigma_{0} \left[ 1+ C
\frac{\alpha_{s}}{\pi} \right] \tau_{H} \frac{d{\cal L}^{gg}}{d\tau_{H}} +
\Delta \sigma_{gg} + \Delta \sigma_{gq} + \Delta \sigma_{q\bar{q}}
\end{equation}
with the renormalization scale in $\alpha_s$ and the factorization
scale of the parton densities to be fixed properly. \s

The coefficient $C(\tau_Q)$ denotes the contributions from the
virtual two--loop corrections regularized by the infrared
singular part of the cross section for real gluon emission. This
coefficient splits into the infrared part $\pi^2$, a
logarithmic term depending on the renormalization scale $\mu$
and a finite $\tau_Q$--dependent piece $c(\tau_Q)$,
\begin{equation}
C(\tau_Q) = \pi^{2}+ c(\tau_Q) + \frac{33-2N_{F}}{6} \log
\frac{\mu^{2}}{m_{H}^{2}}
\label{eq:Cvirt}
\end{equation}
The term $c(\tau_Q)$ can be reduced analytically to a
one--dimensional Feynman--para\-meter integral [see Appendix B] which has
been evaluated numerically \cite{P24}. In the heavy--quark
limit $\tau_Q = m^2_H / 4 m^2_Q \ll 1$ and in the light--quark
limit $\tau_Q \gg 1$, the integrals could be solved
analytically. \s

The (non--singular) hard contributions from gluon radiation in
$gg$ scattering, $gq$ scattering and $q \overline{q}$
annihilation depend on the renormalization scale $\mu$
and the factorization scale $M$ of the parton densities
[Fig.~1b],
\begin{eqnarray}
\Delta \sigma_{gg} & = & \int_{\tau_{H}}^{1} d\tau \frac{d{\cal
L}^{gg}}{d\tau} \times \frac{\alpha_{s}}{\pi} \sigma_{0} \left\{ -
\hat{\tau}
P_{gg} (\hat{\tau}) \log \frac{M^{2}}{\hat{s}} + d_{gg} (\hat{\tau}
,\tau_Q) \right. \non \\
& & \left. \hspace{3.7cm} + 12 \left[ \left(\frac{\log
(1-\hat{\tau})}{1-\hat{\tau}} \right)_+ -\hat{\tau}
[2-\hat{\tau}(1-\hat{\tau})] \log
(1-\hat{\tau}) \right] \right\} \non \\ \non \\
\Delta \sigma_{gq} & = & \int_{\tau_{H}}^{1} d\tau \sum_{q,
\bar{q}} \frac{d{\cal L} ^{gq}}{d\tau} \times \frac{\alpha_{s}}{\pi}
\sigma_{0} \left\{ \hat{\tau} P_{gq}(\hat{\tau})
                   \left[-\frac{1}{2} \log\frac{M^{2}}{\hat{s}} +
\log(1-\hat{\tau}) \right]  +d_{gq} (\hat{\tau}
,\tau_Q) \right\} \non \\ \non \\
\Delta \sigma_{q\bar{q}} & = & \int_{\tau_{H}}^{1} d\tau
\sum_{q} \frac{d{\cal L}^{q\bar{q}}}{d\tau} \times \frac{\alpha_{s}}{\pi}
\sigma_{0}~d_{q\bar q} (\hat{\tau},\tau_Q)
\label{eq:delsig}
\end{eqnarray}
with $\hat{\tau} = \tau_H / \tau$. The renormalization scale enters
through the QCD coupling $\alpha_s (\mu^2)$ in the
radiative corrections and the lowest--order parton
cross section $\sigma_0 [\alpha_s (\mu^2)]$. $P_{gg}$
and $P_{gq}$ are the standard Altarelli--Parisi splitting
functions \cite{P43},
\begin{eqnarray}
P_{gg}(\hat{\tau}) & = & 6 \left\{ \left( \frac{1}{1-\hat{\tau}}
\right)_+ + \frac{1}{\hat{\tau}} -2 +\hat{\tau} (1-\hat{\tau})
\right\} + \frac{33-2N_F}{6}
\delta(1-\hat{\tau}) \nonumber \\
P_{gq}(\hat{\tau}) & = & \frac{4}{3} \frac{1+
(1-\hat{\tau})^2}{\hat{\tau}}
\label{eq:APKernel}
\end{eqnarray}
$F_+$ denotes the usual $+$ distribution such that
$F(\hat{\tau})_+ = F(\hat{\tau}) - \delta (1 - \hat{\tau})
\int_0^1 d\hat{\tau}' F(\hat{\tau}')$. The
coefficients $d_{gg}, d_{gq}$ and $d_{q \ov{q}}$
can be reduced to one--dimensional integrals [Appendix C] which have
been evaluated numerically \cite{P24} for arbitrary quark masses.
They can be solved analytically in the heavy and
light--quark limits. \s

In the heavy--quark limit the coefficients $c(\tau_Q)$ and
$d_{ij} (\hat{\tau},\tau_Q)$ reduce to very simple expressions \cite{P20,P21},
\begin{eqnarray}
\tau_Q = m_H^2/4m_Q^2 \ll 1\,: \hspace{3cm}
c(\tau_Q) & \to & \frac{11}{2} \hspace{6cm} \non \\
d_{gg}(\hat{\tau},\tau_Q) & \to & -\frac{11}{2} (1-\hat{\tau})^3 \non \\
d_{gq}(\hat{\tau},\tau_Q) & \to & -1 + 2 \hat{\tau}
-\frac{\hat{\tau}^2}{3} \non \\
d_{q\bar q}(\hat{\tau},\tau_Q) & \to & \frac{32}{27} (1-\hat{\tau})^3
\end{eqnarray}
The corrections of ${\cal O} (\tau_Q)$ in a systematic Taylor
expansion have been shown to be very small \cite{P41}. In fact,
the leading term provides an excellent approximation
up to the quark threshold $m_H \sim 2 m_Q$. \s

For the sake of completeness we quote the differential
parton cross sections for hard--gluon radiation and quark
scattering in the heavy quark--loop limit \cite{P20,P21,P27,P27A},
\begin{equation}
\frac{d\hat\sigma}{d\hat t} = \frac{G_F \alpha_s^3}{288
\sqrt{2} \pi^2}~H(\hat s, \hat t)
\end{equation}
with
\begin{eqnarray*}
H(gg\to Hg) & = & \frac{3}{2}~
\frac{\hat s^4+\hat t^4+\hat u^4+m_H^8}{\hat{s}^2\hat t\hat u}\\
H(gq\to Hq) & = & -\frac{2}{3}~
\frac{\hat s^2+\hat u^2}{ \hat s \hat t} \\ \\
H(q\bar q\to Hg) & = & \frac{16}{9}~
\frac{\hat t^2+\hat u^2} {\hat{s}^2}
\end{eqnarray*}
The Mandelstam variables $\hat{t}, \hat{u}$ are
the momentum transfer squared from the initial partons
$gg, gq, q\overline{q}$, respectively, to the
Higgs boson in the final state. [The singularities
 for $\hat{t}, \hat{u} \rightarrow 0$ can be
 regularized in $n$ dimensions.] \s

In the opposite limit where the Higgs mass is very
large compared with the top mass, a compact analytic
result can be derived, too:
\begin{eqnarray}
\tau_Q = m_H^2/4m_Q^2 &\gg& 1\,: \hspace{11cm}
\non \\ \non \\
c(\tau_Q) & \to & \frac{5}{36} \log^2 (-4\tau_Q-i\epsilon)
-\frac{4}{3} \log (-4\tau_Q-i\epsilon) \non \\ \non \\
d_{gg}(\hat{\tau},\tau_Q) & \to & -\frac{2}{5} \log(4\tau_Q)
\left\{ 7-7\hat{\tau} +5\hat{\tau}^2 \right\}
- 6 \log (1-\hat{\tau}) \left\{ 1-\hat{\tau} +
\hat{\tau}^2 \right\} \non \\
& & +2\frac{\log \hat{\tau}}{1-\hat{\tau}}
\left\{ 3-6\hat{\tau} -2\hat{\tau}^2
+5\hat{\tau}^3 - 6\hat{\tau}^4 \right\} \non \\ \non \\
d_{gq}(\hat{\tau},\tau_Q) & \to & \frac{2}{3} \left\{ \hat{\tau}^2 -
\left[ 1+(1-\hat{\tau})^2 \right] \left[ \frac{7}{15} \log (4\tau_Q) +
\log\left( \frac{1-\hat{\tau}}{\hat{\tau}} \right)
\right] \right\} \non \\ \non \\
d_{q\bar q}(\hat{\tau},\tau_Q) & \to & 0
\end{eqnarray}
These approximate expressions are valid to leading
and subleading logarithmic accuracy. \s

The final results of our analysis are presented in
Fig.~11 and the subsequent figures for the
LHC energy $\sqrt{s} = 14~\TeV$. [A brief summary is also
given for 10 \TeV.]
                                 They are based on
a top--quark mass of 174 \GeV~\citer{P5,P8A}. If not stated
otherwise, we have adopted the GRV
parameterizations \cite{P44} of the parton densities.
These are defined in the
$\overline{MS}$ scheme\footnote{We may switch
to different schemes by adding the appropriate
 finite shift functions \cite{P42} $f_{ij}$ to the integrals
 $\Delta \sigma_{ij}$, {\it i.e.} substituting $P_{ij} (\hat{\tau})
\log M^2 / \hat{s} \rightarrow P_{ij} (\hat{\tau})\log
M^2 / \hat{s} + f_{ij} (\hat{\tau})$ in eqs.(\ref{eq:delsig}).}.
We have chosen $\alpha_s^{(5)} (m_Z) = 0.117$ of
the $\overline{MS}$ scheme in next--to--leading order.
This corresponds to the average measured QCD coupling for
five quark degrees of freedom \cite{P45} with $\Lambda^{(5)}_
{\overline{MS}} = 214~\MeV$; the standard
matching conditions \cite{P46}
                   are adopted at $\mu = m_t: \alpha_s^{(6)}
 (\mu = m_t | \Lambda_{\ov{MS}}^{(6)}) = \alpha_s^{(5)}
 (\mu = m_t | \Lambda_{\overline{MS}}^{(5)})$ with
 $\Lambda_{\overline{MS}}^{(6)} = 0.413~\Lambda _{\overline{MS}}^{(5)}$.
The GRV fits are based on a somewhat smaller value of $\alpha_s$.
This introduces a slight inconsistency into the numerical
evaluation of the cross section which we allow for since,
on the other hand, the basic parton cross section is quadratic
in $\alpha_s$ and thus depends strongly on the choice of the
QCD coupling.
In order to correct the difference in the $\Lambda_{\overline{MS}}$
values, the factorization scale $M$ at which the parton densities are
evaluated, has been changed to adjust appropriately the ratio
$M^2/\Lambda^2_{\overline{MS}}$ which enters in the structure
functions\footnote{The dependence of the cross
section on the factorization scale is very small.}.
The cross section is sensitive to gluon
 and quark densities down to $x$ values of order
 $10^{-2}$ to $10^{-3}$, so that subtle non--linear
 effects in the evolution at small $x$ need not be
 taken into account yet. \s

We introduce $K$ factors in the standard way,
\begin{eqnarray}
K_{tot} = \frac{\sigma_{HO}}{\sigma_{LO}}
\end{eqnarray}
The cross sections $\sigma_{HO}$ in next--to--leading order
are normalized to the cross sections $\sigma_{LO}$,
evaluated consistently for parton densities and $\alpha_s$
in leading order; the QCD NLO and LO couplings are
taken from the GRV parameterizations
of the structure functions.
The $K$ factor can be broken down to
several characteristic components. $K_{virt}$ accounts
for the regularized virtual corrections, corresponding
to the coefficient $C$; $K_{AB}$ [$A,B=g, q,
\bar{q}$] for the real corrections as defined in eqs.(\ref{eq:delsig}).
These $K$ factors are shown for LHC
energies in Fig.~11 as a function of the Higgs boson mass.
For both the renormalization and the factorization scales,
$\mu = M = m_H$ has been chosen. Apparently $K_{virt}$
and $K_{gg}$ are of the same size and of order 50$\%$ while
$K_{gq}$ and $K_{q\overline{q}}$ are quite small.
[Note that $(K_{virt} + \Sigma K_{AB})$ differs
from $(K_{tot} - 1)$ since the cross sections
$\sigma_0$ are evaluated with different NLO and LO
$\alpha_s$ values in the numerator and denominator.]
Apart from the threshold region for Higgs decays into
$t \overline{t}$ pairs, $K_{tot}$ is insensitive to
the Higgs mass. \s

The absolute magnitude of the correction is positive
and large, increasing the cross section for Higgs
production at the LHC significantly by a factor
of about 1.5 to 1.7. Comparing the exact numerical results
with the analytic expressions in the heavy--quark limit, it
turns out that these asymptotic solutions provide
an excellent approximation even for Higgs masses
above the top--decay threshold. For Higgs masses
below $\sim$ 700 \GeV, the deviations of the QCD
corrections from the asymptotic approximation are
less than 10\%. \s

There are two sources of uncertainties in the theoretical
prediction of the Higgs cross section, the variation
of the cross section with different parametrizations of the
parton densities and the unknown next--to--next--to--leading
corrections. Since all mass scales, the Higgs mass as well as
the loop--quark mass, are very large, the notorious
uncertainties from higher--twist effects can safely be
assumed absent. \s

One of the main uncertainties in the prediction of the Higgs
production cross section is due to the gluon density.
This distribution can only indirectly be extracted through
order $\alpha_s$ effects from deep--inelastic
lepton--nucleon scattering, or through complicated analyses
of final states in lepton--nucleon and hadron--hadron
scattering. Adopting a set of representative parton
distributions \cite{P44,P47,P47A} which are up--to--date fits to all
available experimental data, we find a variation
of about 7\% between the maximum and minimum values
of the cross section for Higgs masses above $\sim $
100 \GeV, Fig.~12a. This uncertainty will be reduced in
the near future when the deep--inelastic
electron/positron--nucleon scattering experiments at
HERA will have reached the anticipated level of accuracy. \s

The [unphysical] variation of the cross section with the
renormalization and factorization scales is reduced by
including the next--to--leading order corrections. This is
demonstrated in Fig.~13 for two typical values
of the Higgs mass, $m_H = 150~\GeV$ and $m_H = 500~\GeV.$
The renormalization/factorization scale $\mu = M$
is varied as $\mu = \rho m_H$ for $\rho$ between 1/2 and 2.
The ratio of the cross sections is reduced from 1.62
in leading order to 1.32 in next--to--leading order for
 $m_H = 500~\GeV$.
While for small Higgs masses the variation with $\mu$ for
$\rho \sim 1$ is already small at the LO level, the
improvement by the NLO corrections is significant  at
the NLO level for large Higgs masses.
However, the figures indicate that further improvements
are required since the $\mu$ dependence of the cross
section is still monotonic in the
parameter range set by the scale of order $m_H$.
These uncertainties associated with
higher--order corrections appear to be less than
about 15\% however. \s

If the total energy is reduced from $\sqrt{s}$ = 14 \TeV~to
10 \TeV~the production cross section for the
 $\SM$ Higgs boson decreases by a little less than a
factor 2 for small Higgs masses and a little more than 2
for large Higgs masses, Fig.~12b. The $K$ factors agree within
less than $\sim 5\%$ for the two energies.

\vspace*{0.3cm}

\section{The Neutral $\SUSY$ Higgs Particles}

\subsection{The Basic Set--Up}

Supersymmetric theories are very attractive extensions of the Standard Model.
At low energies they provide a theoretical framework in
which the hierarchy problem in the Higgs sector is solved while retaining
Higgs bosons with moderate masses as elementary particles in the context of
the high mass scales demanded by grand unification.  The minimal
supersymmetric extension of the Standard Model ($\MSSM$) \cite{P15} may serve
as a useful guideline in this domain \cite{X8}.  This point is underlined by
the fact that the model led to a prediction of the electroweak mixing angle
\cite{P24B} that is in striking agreement with present high--precision
measurements of $\sin^2\theta_W$ \cite{A2}.  Although some of the phenomena
will be specific to this
minimal version, the general pattern will nevertheless be characteristic to
more general extensions \cite{P18,P48} so that the analyses can be considered
as
representative for a wide class of $\SUSY$ models.  \s

Supersymmetry requires the existence of at least two isodoublet scalar fields
$\Phi_{1}$ and $\Phi_{2}$, thus extending the physical spectrum of scalar
particles  to five \cite{P24A}.
The $\MSSM$ is restricted to this minimal extension. The
field $\Phi_{2}$ [with vacuum expectation value $v_{2}$] couples only to
up--type quarks while $\Phi_{1}$ [with vacuum expectation value $v_{1}$]
couples to down--type quarks and charged leptons. The physical Higgs bosons
introduced by this extension are of the following type: two ${\cal CP}$--even
neutral bosons $h^0$ and $H^0$ [where $h^0$ will be the lightest particle],
a ${\cal CP}$--odd neutral boson $A^0$ [usually called pseudoscalar] and
two charged Higgs
bosons $H^{\pm}$. \s

Besides the four masses $m_{h^0}$, $m_{H^0}$, $m_{A^0}$ and $m_{H^\pm}$,
two additional parameters define the properties of the scalar particles and
their interactions with gauge bosons and fermions: the mixing angle
$\beta$, related to the ratio of the two
vacuum expectation values $\tb = v_2/v_1$, and the mixing angle
                  $\alpha$ in the
neutral ${\cal CP}$--even sector. Supersymmetry gives rise to several
relations among these parameters and, in fact, only two of them are
independent.  These relations impose a strong hierarchical structure on the
mass spectrum $ [ m_{h^0}<m_Z , m_{A^0} < m_{H^0}$ and
                                        $m_W <m_{H^\pm}]$  which
                                however is
broken by radiative corrections \cite{P17,P26A} due to the large
                                       top quark mass. The
parameter $\tg \beta$ will in general be assumed in the range $1 < \tg
\beta < m_t/m_b$ $[ \pi/4 < \beta < \pi/2] $, consistent with
the restrictions
that follow from interpreting the $\MSSM$ as the
                                                        low energy limit of a
supergravity model. \s

The $\MSSM$ Higgs sector is generally parameterized by the mass
$m_{A^0}$ of the pseudoscalar Higgs boson and $\tb$.
                           Once these two parameters [as well as the top quark
mass and the associated squark masses which enter through radiative
corrections] are specified, all other masses and the mixing angle $\alpha$ can
be predicted. To discuss the radiative corrections we shall neglect, for
the sake of simplicity, non--leading effects due to non--zero values of the
supersymmetric Higgs mass parameter $\mu$ and of the parameters $A_t$ and
$A_b$ in the soft symmetry breaking interaction. The radiative corrections are
then determined by the parameter $\epsilon$ which grows as the fourth power of
the top quark mass $m_t$ and logarithmically with the squark mass $M_S$,
 \begin{eqnarray}
 \epsilon = \frac{3 \alpha}{2 \pi} \frac{1}{s_W^2 c_W^2}
 \frac{1}{\sin^2 \beta} \frac{m_t^4}{m_Z^2} \log \left( 1+
 \frac{M_S^2}{m_t^2} \right)
\label{eq:epsusy}
 \end{eqnarray}
with $s_W^2=1-c_W^2 = \sin^2 \theta_W$.  [The main part of the two--loop
effects can be incorporated by using the running $\overline{MS}$ top mass
evaluated at the pole mass \cite{P61A}.]  \s

                                 These corrections are positive
and they shift the mass of the light neutral Higgs boson $h^0$
                                                             upward with
increasing top mass. The variation of the
upper limit on $m_{h^0}$ with
the top quark mass is shown in Fig.~14a for $M_S=$ 1 \TeV~and two
representative values of
$\tb = 1.5$ and 30. While the dashed lines correspond to the
leading radiative corrections in eq.(\ref{eq:epsusy}),
\begin{equation}
m^2_{h^0} \le m_Z^2 \cos^2 2\beta + \epsilon \sin^2\beta
\label{eq:hbound}
\end{equation}
\nn the solid lines correspond to the Higgs mass parameter
$\mu = - 200,0, + 200~\GeV$ and the Yukawa parameters
$A_t = A_b = 1~\TeV$. The upper bound on $m_{h^0}$ is
shifted from the tree--level value $m_Z$ up to
$\sim$ 140 \GeV~for $m_t = 174~\GeV$. \s

Taking $m_{A^0}$ and $\tb$ as the base parameters, the mass of the lightest
scalar state $h^0$ is given to leading order by
\begin{eqnarray}
m^2_{h^0} & = & \frac{1}{2} \left[ m_{A^0}^2 + m_Z^2 + \epsilon \right.
\non \\
& & \left. - \sqrt{(m_{A^0}^2+m_Z^2+\epsilon)^2
-4 m_{A^0}^2m_Z^2 \cos^2 2\beta
-4\epsilon (m_{A^0}^2 \sin^2\beta + m_Z^2 \cos^2\beta)} \right]
\label{eq:hmass}
\end{eqnarray}
\nn The masses of the heavy neutral and charged Higgs bosons
follow from the sum rules
\begin{eqnarray}
m_{H^0}^2 & = & m_{A^0}^2 + m_Z^2 - m_{h^0}^2 + \epsilon \non \\
m_{H^\pm}^2 & = & m_{A^0}^2 + m_W^2
\end{eqnarray}
\nn In the subsequent discussion we will assume for definiteness
that $m_t = 174~\GeV$, $M_S = 1~\TeV$ and $\mu=A_t=A_b=0$.
For the two representative values of
     $\tb$ introduced above, the masses $m_{h^0},
m_{H^0}$ and $m_{H^\pm}$ are displayed in Figs.~14b--d as
a function of $m_{A^0}$. [The dependence of
the masses on the parameters $\mu, A_t, A_b$ is weak and
the mass shifts are limited by a few \GeV~\cite{X6A}.] \s

The mixing parameter $\alpha$ is determined by $\tb$ and
the Higgs mass $m_{A^0}$,
\begin{equation}
\tg 2 \alpha = \tg 2\beta \frac{m_{A^0}^2 + m_Z^2}{m_{A^0}^2 - m_Z^2 +
\epsilon/\cos 2\beta} \ \ \ \ \mbox{with} \ \ \  -\frac{\pi}{2} < \alpha <0
\end{equation}

\vspace*{2mm}

The couplings of the various neutral Higgs bosons to fermions and gauge
bosons depend on the angles $\alpha$ and $\beta$. Normalized
to the $\SM$ Higgs couplings, they are summarized in Table \ref{tb:hcoup}.
The pseudoscalar particle $A^0$
             has no tree level couplings to gauge bosons, and its couplings to
down (up)--type fermions are (inversely) proportional to $\tb$.

\begin{table}[hbt]
\renewcommand{\arraystretch}{1.5}
\begin{center}
\begin{tabular}{|lc||ccc|} \hline
\multicolumn{2}{|c||}{$\Phi$} & $g_{\Phi \bar{u}u}$
& $g_{\Phi \bar{d}d}$ &  $g_{\Phi VV}$ \\ \hline \hline
\SM~ & $H$ & 1 & 1 & 1 \\ \hline
\MSSM~ & $h^0$ & $\cos\alpha/\sin\beta$ & $-\sin\alpha/\cos\beta$ &
$\sin(\beta-\alpha)$ \\
& $H^0$ & $\sin\alpha/\sin\beta$ & $\cos\alpha/\cos\beta$ &
$\cos(\beta-\alpha)$ \\
& $A^0$ & $ 1/\tg\beta$ & $\tg\beta$ & 0 \\ \hline
\end{tabular}
\renewcommand{\arraystretch}{1.2}
\caption{\label{tb:hcoup}
{\small Higgs couplings in the $\MSSM$ to fermions and gauge bosons
relative to $\SM$ couplings.}}
\end{center}
\end{table}

\vspace*{-3mm}

Typical numerical values of these couplings are shown in Fig.~15 as a function
of $m_{A^0}$ and for two values of $\tb$.  The dependence on the parameters
$\mu$ and $A_t,A_b$ is very weak and the leading radiative corrections provide
an excellent approximation \cite{X6A}.  There is in general a strong
dependence on the input parameters $\tb$ and $m_{A^0}$.  The couplings to down
(up)--type fermions are enhanced (suppressed) compared to the $\SM$ Higgs
couplings.  If $m_{A^0}$ is large, the couplings of $h^0$ to fermions and
gauge bosons
are $\SM$ like.  It is therefore very difficult to distinguish the Higgs
sector of the $\MSSM$ from that of the $\SM$, if all Higgs bosons, except the
lightest neutral Higgs boson, are very heavy.  \s

Apart from cascade decays in some corners of the $\SUSY$ parameter space, the
main decay modes of the neutral Higgs particles are in general $b \bar{b}$
decays [$\sim 90 \%$] and $\tau^+ \tau^-$ decays [$\sim 10 \%$], and top
decays above threshold.  The
branching ratios for all the dominant decay modes are shown in Fig.~16.  The
gold--plated $ZZ$ decays of the $\SM$ Higgs particle above 140 \GeV~play only
a minor r\^ole in the $\SUSY$ Higgs sector --- and in large parts of the
parameter space their r\^ole is even negligible.  The total widths of the
states remain small, ${\cal O}(1~\GeV)$, anywhere in the intermediate mass
range and they do not exceed a few tens of \GeV~even for Higgs masses of the
order of 1 \TeV,  Fig.~17.  \s

In addition to the conventional decays into $\SM$
particles, the Higgs particles may also decay into chargino and
neutralino pairs \cite{X6,X6A}.
Depending on the details of the $\SUSY$ parameters, the branching ratios for
decays into these channels can add up to a few tens of percent;
invisible LSP (lightest neutralino) decays, in particular, can even
dominate in some
domains of the $\MSSM$ parameter space. When
kinematically allowed, the Higgs particles also decay into squarks and
sleptons, with generally small branching ratios, though.
For the present experimental bounds on non--colored and colored
supersymmetric particles, see Refs.~\cite{P49} and \cite{P52},
respectively. \s

The neutral Higgs particles will be searched for mainly in the
decay channels $\tau^+ \tau^-$ and $\gamma \gamma$ at the LHC
\cite{P14B,P14A}. Large QCD backgrounds render the analysis of the dominating
$b \ov{b}$ final states very difficult. Nonetheless, detailed
feasibility studies have demonstrated that the $b \ov{b}$
decay channel \cite{A5} may be accessible in associated $Wh^0$ and
$t \ov{t} / b \ov{b} h^0$ events if a set of strong detector
requirements is met \cite{P14}.

\subsection{The Two--Photon Decay Widths}
Similarly to the Standard Model Higgs boson, the precise prediction
of the $\gamma \gamma$ widths of the $\SUSY$ Higgs particles
is motivated by several points. This rare decay mode provides
the most important signature for the search of the light
Higgs bosons at hadron colliders. The values of the coupling
constants are affected by the charged particle loops of the
entire $\SUSY$ spectrum with masses far exceeding the
light Higgs mass. The effect however is small in general
for heavy $\SUSY$ particles since the main component of
their masses is not generated by the Higgs mechanism so
that these particles decouple asymptotically. \s

The $\gamma \gamma$ coupling to Higgs bosons in supersymmetric
theories is mediated by charged heavy particle loops built up
by $W$ bosons, standard fermions $f$, charged Higgs
bosons $H^\pm$, charginos $\tilde{c}$ and sfermions
$\tilde{f}$ in the scalar cases $h^0, H^0$, and standard
fermions and charginos [in the absence of sfermion mixing]
in the pseudoscalar case $A^0$.
Denoting the amplitudes by $A_f$ etc., the
$\gamma \gamma$ decay rates are given\footnote{The scalar
particles $h^0, H^0$ will generically be denoted by ${\cal H}^0$,
and all the neutral Higgs particles by $\Phi$.}
by
\begin{eqnarray}
\Gamma({\cal H}^0 \rightarrow\gamma\gamma) & = &
\frac{G_F\alpha^2m_{\cal H}^3}{128\sqrt{2}\pi^3}
\left| \sum_f N_c e_f^2 g_f^{\cal H} A_f^{\cal H} +
g_W^{\cal H} A_W^{\cal H} + g_{H^\pm}^{\cal H} A_{H^\pm}^{\cal H} \right.
\non \\
& & \left. \hspace{3.2cm} + \sum_{\tilde c} g_{\tilde c}^{\cal H}
A_{\tilde c}^{\cal H} +
\sum_{\tilde f} N_c e_{\tilde f}^2 g_{\tilde f}^{\cal H} A_{\tilde f}^{\cal H}
\right|^2
\end{eqnarray}
and
\begin{equation}
\Gamma(A^0 \rightarrow\gamma\gamma)=
\frac{G_F\alpha^2m_A^3}{32\sqrt{2}\pi^3}
\left| \sum_f N_c e_f^2 g_f^A A_f^A
+ \sum_{\tilde c} g_{\tilde c}^A A_{\tilde c}^A
\right|^2
\end{equation}
The spin 1, spin 1/2 and spin 0 amplitudes read to lowest
order for the scalar Higgs bosons
\begin{eqnarray}
A_1^{\cal H}     & = & - [2\tau^2+3\tau+3 (2\tau-1)f(\tau)]/ \tau^2 \non \\
A_{1/2}^{\cal H} & = & 2 [\tau +(\tau-1)f(\tau)]/\tau^2 \nonumber \\
A_0^{\cal H}     & = & - [\tau - f(\tau)]/\tau^2
\end{eqnarray}
and for the pseudoscalar Higgs boson
\begin{eqnarray}
A_{1/2}^A & = & f(\tau)/\tau
\end{eqnarray}

\begin{table}[hbt]
\renewcommand{\arraystretch}{1.6}
\begin{center}
\begin{tabular}{|lc||cc|} \hline
\multicolumn{2}{|c||}{$\Phi$} & $H^c$ & $\tilde c_i$ \\
\hline \hline
\SM~ & $H$ & 0 & 0 \\ \hline
\MSSM~ & $h^0$ & $\frac{m_W^2}{m_{H^c}^2} \left[ \sin(\beta-\alpha)+\frac{\cos
2\beta \sin (\beta+\alpha)}{2\cos^2 \theta_W} \right]$
& $2\frac{m_W}{m_{\tilde c_i}} (S_{ii} \cos\alpha - Q_{ii} \sin\alpha)$ \\
& $H^0$ & $\frac{m_W^2}{m_{H^c}^2} \left[ \cos(\beta-\alpha)-\frac{\cos
2\beta \cos (\beta+\alpha)}{2\cos^2 \theta_W} \right]$
& $2\frac{m_W}{m_{\tilde c_i}} (S_{ii} \sin\alpha + Q_{ii} \cos\alpha)$ \\
& $A^0$ & 0 & $2\frac{m_W}{m_{\tilde c_i}} (-S_{ii} \cos\beta - Q_{ii}
\sin\beta)$ \\[0.2cm] \hline
\end{tabular} \\[0.3cm]

\begin{tabular}{|lc||c|} \hline
\multicolumn{2}{|c||}{$\Phi$} & $\tilde f_{L,R}$ \\
\hline \hline
\SM~ & $H$ & 0 \\ \hline
\MSSM~ & $h^0$ & $\frac{m_f^2}{m_{\tilde f}^2} g_f^h \mp
\frac{m_Z^2}{m_{\tilde f}^2} (I_3^f - e_f \sin^2\theta_W) \sin
(\alpha+\beta)$ \\
& $H^0$ & $\frac{m_f^2}{m_{\tilde f}^2} g_f^H \pm
\frac{m_Z^2}{m_{\tilde f}^2} (I_3^f - e_f \sin^2\theta_W) \cos
(\alpha+\beta)$ \\
& $A^0$ & 0 \\ \hline
\end{tabular}
\renewcommand{\arraystretch}{1.2}
\caption[ ]{\label{tb:hcoup2}
{\small Higgs couplings in the $\MSSM$ to charged
Higgs bosons, charginos and sfermions relative to $\SM$ couplings.
$Q_{ii}$ and $S_{ii}$ $(i=1,2)$ are related to the mixing angles between
the charginos $\tilde c_1$ and $\tilde c_2$, Ref.\cite{X8}.}}
\end{center}
\end{table}

\nn As usual, the scaling variable is defined as $\tau = m^2_{\Phi}/4m^2_i$
with $m_i$ denoting the loop mass.  The universal scaling function $f(\tau)$
is the same as in eq.(\ref{eq:ftau}).  The coefficients $g^{\Phi}_i$ denote
the couplings of the Higgs bosons to $W$ bosons, top and bottom quarks given
in Table \ref{tb:hcoup} and the couplings to sfermions and charginos which are
recollected for the sake of convenience in Table \ref{tb:hcoup2} in the
absence of sfermion mixing.  [Including mixing effects in the scalar squark
sector due to the soft parameters $A_t,~A_b$ and $\mu$ does not change the
production cross sections and photonic decay widths of the \SUSY~Higgs bosons
in most of the parameter space, except in small regions where they play a
significant r\^ole and lead to an enhancement of the signal \cite{P53}.] \s

Since the contributions of the squark loops are strongly
suppressed compared to $t, b$ loops, we shall restrict
the discussion of the QCD corrections to the standard
quark loops. These corrections will be parameterized again as
\begin{equation}
A_Q = A_Q^{LO} \left[ 1+C \frac{\alpha_s}{\pi} \right]
\end{equation}
The coefficient $C$ depends on $\tau = m^2_{\Phi} / 4 m^2_Q (\mu_Q^2)$,
 where the running mass $m_Q (\mu_Q^2)$ is defined at the
 renormalization point $\mu_Q$,
 \begin{equation}
C = c_1 [m_Q(\mu_Q^2)] + c_2 [m_Q(\mu_Q^2)] \log \frac{\mu_Q^2}{m_Q^2}
\label{eq:3.c1c2}
 \end{equation}
The renormalization point is taken to be $\mu_Q = m_{\Phi} / 2 $;  this value
is related to the pole mass by the QCD formula noted in eq.(\ref{eq:mrun}).
The lowest order amplitude $A^{LO}_Q$ must be evaluated for the same mass
value $m_Q (\mu_Q^2=[m_{\Phi} / 2]^2)$.  The choice $\mu_Q = m_{\Phi} / 2$ of
the renormalization point ensures, {\it a posteriori}, a behavior of the
$\gamma \gamma$ couplings which is well controlled for Higgs masses much
larger than the quark mass.  The QCD coupling $\alpha_s$ is evaluated at $\mu
= m_{\Phi}$ for $\Lambda_{\overline{MS}}^{(5)} = 214~\MeV$.  \s

To regularize the pseudoscalar amplitude involving
the $\gamma_5$ coupling, we have adopted the 't Hooft--Veltman
prescription \cite{P54}. A technical remark ought to be added on a subtle
problem related to this implementation of $\gamma_5$ which
reproduces the axial--vector anomaly to lowest order
automatically. The multiplicative renormalization factor of the
scalar $(Q \ov{Q})$ current is given by $Z_{{\cal H}QQ} =
1 - Z_2 Z_m$ where $Z_2, Z_m$ are the
wave--function and mass renormalization factors, respectively.
 To ensure the chiral--symmetry relation
$\Gamma_5 (p', p) \to \gamma_5 \Gamma (p', p)$ in the
limit $m_Q \to 0$ for the fermionic matrix element of the
pseudoscalar and scalar currents, the renormalization factor
of the pseudoscalar current has to be chosen \cite{P55} as
\begin{eqnarray}
Z_{AQQ} = Z_{{\cal H}QQ} + \frac{8}{3}~\frac{\alpha_s}{\pi}
\end{eqnarray}
The additional term, supplementing the naive expectation, is due to
spurious anomalous contributions that must be subtracted
by hand. \s

\vspace*{2mm}

\noindent
\underline{\it The Limit of Large Loop Masses}  \s

\vspace*{2mm}

\nn For large $m_Q$, the coefficient $c_2$ in eq.(\ref{eq:3.c1c2}) is of order
$1/m_Q^2$ and approaches zero for the scalar and pseudoscalar Higgs bosons.
It has been shown before that $c_1$ approaches $(-1)$ for scalar Higgs bosons;
for the pseudoscalar Higgs particle $c_1$ vanishes asymptotically, i.e.
\begin{eqnarray}
m_Q \to \infty \,: \hspace{0.5cm}
c_1^{\cal H} & \to & -1 \non \\
c_1^A        & \to &  0
\end{eqnarray}
This result for the pseudoscalar Higgs boson can also
be derived from
the non--renormali\-zation of the anomaly of the axial--vector
current.
 In the same way in which the ${\cal H}\gamma\gamma$ coupling
in the local limit can be related to the anomaly of the trace
of the energy--momentum tensor, we can derive the $A^0 \gamma
 \gamma$ coupling from the anomaly of the axial--vector current
\cite{P56},
\begin{equation}
\partial_\mu j_\mu^5 = 2m_Q \overline{Q} i\gamma_5 Q +
N_c e_Q^2\frac{\alpha}{4\pi} F_{\mu\nu} \widetilde{F}_{\mu\nu}
\label{eq:anomaly}
\end{equation}
with $\widetilde{F}_{\mu\nu} = \epsilon_{\mu\nu\alpha\beta}
        F_{\alpha\beta}$ denoting the dual field strength tensor.
Since, as familiar from the Sutherland--Veltman paradox,
the matrix element $\langle \gamma \gamma | \partial_\mu j^\mu_5 |
0 \rangle$ of the divergence of the axial-vector current vanishes
for zero photon energy, the matrix element
$\langle \gamma \gamma | m_Q \ov{Q} i \gamma_5 Q|0\rangle$ of the
Higgs source can be linked directly to the anomalous term in
eq.(\ref{eq:anomaly}). It is well--known that the anomaly is not renormalized
if the QCD strong interactions are switched on \cite{P56}. As a result, the
effective $A^0 \gamma \gamma$ Lagrangian
\begin{equation}
{\cal L}_{eff} (A^0 \gamma\gamma) = N_c e_Q^2 \frac{\alpha}{8\pi}
\left(\sqrt{2}
G_F \right)^{1/2} F_{\mu\nu} \widetilde{F}_{\mu\nu} A^0
\end{equation}
\nn is valid to all orders of perturbation theory in $\alpha_s$ in the
limit $m^2_{A^0} \ll 4 m^2_Q$. \s

\vspace*{2mm}

\noindent
\underline{\it The Limit of Small Loop Masses} \s

\vspace*{2mm}

\nn
Also in the opposite limit of small quark--loop masses compared with the
Higgs masses, the ${\cal H}^0 \gamma \gamma$ and $A^0 \gamma \gamma$
couplings can be calculated analytically. This limit
is useful in practice for large $\tb$ values
where the $b$ quark coupling to the heavy Higgs bosons
$H^0$ and $A^0$ is strongly enhanced. As anticipated theoretically,
 the leading and subleading logarithmic terms are chirally invariant and
we obtain the same QCD correction in this limit for the
 scalar and pseudoscalar couplings,
\begin{equation}
m_Q(\mu_Q^2) \to 0\,: \hspace{0.5cm} C^{{\cal H},A} \to -\frac{1}{18}
\log^2 (-4\tau - i\epsilon) - \frac{2}{3}
\log (-4\tau -i\epsilon) + 2\log\frac{\mu_Q^2}{m_Q^2}
\end{equation}
The finite non--logarithmic contributions to $C$ may be
different in the scalar and pseudoscalar cases. \s

The amplitudes $C^{\cal H}$ for scalar loops and $C^A$ for
pseudoscalar loops are shown in Fig.~18 as a function
of $\tau$ \cite{P22}. The coefficients are real below the quark
threshold $\tau < 1$, and complex above. Very close
to the threshold, within a margin of a few \GeV, the
present perturbative analysis\footnote{By choosing
the renormalization point $\mu_Q = m_{\Phi} / 2$ the
perturbative threshold $E_{th} = 2m_Q (m^2_Q)$
coincides with the on--mass shell value proper. A shift
between $m_{\Phi} / 2$ and $m_{\Phi}$, for instance,
affects the widths very little away from the threshold.}
can not be applied. [It may account to some
 extent for resonance effects in a global way.]
Since $Q \ov{Q}$ pairs cannot form $0^{++}$ states
at the threshold, $\Im mC_{\cal H}$ vanishes there; $\Re eC_{\cal H}$
develops a maximum very close to the threshold. By
contrast, since $Q \ov{Q}$ pairs do form $0^{-+}$
states, the imaginary part $\Im mC_A$ develops a step
that is built up by the Coulombic gluon exchange [familiar
from the Sommerfeld singularity of the QCD correction
to $Q \ov{Q}$ production in $e^+e^-$ annihilation];
$\Re eC_A$ is singular at the threshold. \s

The singular behavior of the $A^0 \gamma \gamma$ coupling
demands a more careful analysis at the quark threshold
\cite{P29}. The form factor is given to lowest order
near the threshold by
\begin{equation}
A_Q^{A,LO} (\tau_Q) =
               f(\tau_Q) / \tau_Q \to \frac{\pi^2}{4} + i\pi\beta
\hspace{0.5cm} \mbox{~for~} \tau_Q \to 1
\end{equation}
Where $\beta = \sqrt{1 - \tau^{-1}_Q}$ is the quark velocity above
the threshold. The QCD corrections to the imaginary part can be found
by attaching the Sommerfeld rescattering correction \cite{P57}
\begin{equation}
C_{Coul} = \frac{Z}{1-e^{-Z}} \approx 1 + \frac{1}{2} Z
\hspace{0.5cm} \mbox{~for~} Z = \frac{4 \pi \alpha_s}
{3 \beta}
\end{equation}
which corresponds to the exchange of a ladder of Coulombic gluon
quanta between the slowly moving quarks. The QCD corrected
imaginary part of the $A^0 \gamma \gamma$ coupling may thus be
written
\begin{equation}
\Im m~A_Q^A =
\pi \beta C_{Coul} = \pi \beta + \frac{2}{3} \pi^2 \alpha_s
\label{eq:Imacoul}
\end{equation}
approaching a non--zero value at threshold. The real part can be derived from
a once--subtracted dispersion relation so that near the
threshold
\begin{equation}
 A_Q^A \to A_Q^{A,LO} + \frac{2 \pi \alpha_s}{3}
\left[ -\log(\tau_Q-1) + i\pi + const \right]
\end{equation}
The smooth constant term needs not be fixed if we analyze only
the singular behavior. For the QCD correction $C^A$ near the
threshold we therefore obtain the simple relations,
\begin{eqnarray}
\tau_Q\to 1: \hspace{0.5cm} \Re e~C^A & \to &
                                         -\frac{8}{3}  \log(\tau_Q
-1) + const \non \\
\Im m~C^A & \to & +\frac{8}{3} \pi \approx 8.38
\end{eqnarray}
The absolute size of the imaginary part and the logarithmic
singularity of the real part are in agreement with the numerical
analysis presented in Fig.~18b. \s

In Figs.~19a,b the QCD corrected $\gamma \gamma$ widths
for the $h^0, H^0, A^0$ Higgs bosons are
displayed, taking into account
 only quark and $W$ boson loops for two values
$\tb = 1.5$ and $\tb = 30$. While in the
first case top loops give a significant contribution, bottom
loops are the dominant component for large $\tb$.
The overall QCD corrections are shown in Figs.~19c,d.
The corrections to the widths are small,
$\sim {\cal O} (\alpha_s/\pi)$ everywhere.
[Artificially large values of $\delta$ occur only for
specific large Higgs masses when the lowest order
amplitudes vanish accidentally as a consequence
of the destructive interference between $W$ and
quark--loop amplitudes, see also \cite{P33}.] Thus, the
QCD corrections are well under control across the
physically interesting mass range if the
running of the quark masses is properly taken into
account.

\subsection{The Gluonic Decay Widths}
The gluonic decays of the Higgs bosons
\begin{displaymath}
h^0,H^0,A^0 \to gg
\end{displaymath}
are mediated by quark and squark triangle loops. In the same
    notation as in the preceding section
we find for the widths in lowest order
\begin{eqnarray}
\Gamma(h^0\rightarrow gg) & = & \frac{G_F \alpha_s^2}{36\sqrt{2}\pi^3} m_h^3
\left| \frac{3}{4}\sum_Q g_Q^h A^h_Q + \frac{3}{4}\sum_{\widetilde{Q}}
g_{\widetilde{Q}}^h A_{\widetilde{Q}}^h \right|^2 \non \\
\Gamma(H^0\rightarrow gg) & = & \frac{G_F \alpha_s^2}{36\sqrt{2}\pi^3} m_H^3
\left| \frac{3}{4} \sum_Q g_Q^H A^H_Q + \frac{3}{4} \sum_{\widetilde{Q}}
g_{\widetilde{Q}}^H A_{\widetilde{Q}}^H \right|^2 \non \\
\Gamma(A^0\rightarrow gg) & = & \frac{G_F \alpha_s^2}{16\sqrt{2}\pi^3} m_A^3
\left| \sum_Q g_Q^A A^A_Q \right|^2
\end{eqnarray}
Since the contribution of heavy squark loops is small, we will neglect these
effects in the following discussion and we will focus on the dominant quark
contributions. \s

The QCD corrections to the gluonic decay widths are large. Besides the virtual
corrections, the widths are affected by three--gluon and gluon plus light
quark--antiquark final states,
\begin{equation}
h^0,H^0,A^0 \to ggg \mbox{~and~} gq\bar q
\end{equation}
Proceeding in the same way as for the Standard Model, the result can be
written in the form [$\Phi = h^0, H^0, A^0$]
\begin{equation}
\Gamma(\Phi \rightarrow gg(g),~gq\bar q) =
\Gamma_{LO}(\Phi \rightarrow gg) \left[ 1 + E_\Phi(\tau_Q)
\frac{\alpha_s}{\pi} \right]
\end{equation}
with
\begin{eqnarray}
E_{\cal H}(\tau_Q) & = & \frac{95}{4} - \frac{7}{6} N_F
+ \frac{33-2N_F}{6}\ \log \frac{\mu^2}{m_{\cal H}^2} + \Delta E_{\cal H}
\non \\ \non \\
E_A(\tau_Q)        & = & \frac{97}{4} - \frac{7}{6} N_F
+ \frac{33-2N_F}{6}\ \log \frac{\mu^2}{m_A^2} + \Delta E_A
\end{eqnarray}

In the limit of large loop masses, a contribution 11/2 to the coefficients for
scalar states is related to the effective Lagrangian after the heavy quarks
are integrated out;  the remaining part is associated with the rescattering
and splitting corrections.  As a result of the non--renormalization of the
axial anomaly, the coefficient for the pseudoscalar state is entirely
due to the rescattering and splitting corrections.  The corrections $\Delta
E_\Phi$ are displayed in Figs.~20a,b as functions of the corresponding Higgs
masses within their relevant mass ranges for $\tb = 1.5$ and 30.  Due to the
bottom contribution, the deviations from the heavy quark--loop limit are
significantly larger than in the \SM~case, thus rendering this limit useful
only for $\tb$ close to unity.  In Fig.~21 the gluonic decay widths
including the QCD radiative corrections are presented for $\tb = 1.5$ and 30.
They are enhanced by about 50\% to 70\% as a result of the large QCD
corrections.  In a margin of a few \GeV~near the threshold [$m_A \approx
2m_t$] the perturbative result of the pseudoscalar decay width is not valid
due to the Coulomb singularity in analogy to the photonic decay $A^0\to \gamma
\gamma$.  The final branching ratios of all decay processes in the \MSSM~are
shown in Fig.~16.  For the light Higgs particle $h^0$ the gluonic decay mode
is significant only for $h^0$ masses close to the maximal value, where $h^0$
has \SM~like couplings.  For $H^0$ the gluon decay mode is significant only
slightly below the top--antitop threshold and for small values of $\tb$ where
the coupling to top quarks is sufficiently large.  For the pseudoscalar Higgs
boson $A^0$, the gluonic decay mode is important for small values of $\tb$ and
below the top--antitop threshold, where it can reach a branching fraction of
$\sim 20\%$.  \s

In the limit of large quark masses, the Higgs--gluon--gluon coupling can be
described by gauge--invariant effective Lagrangians,
\begin{eqnarray}
{\cal L}_{{\cal H}^0 gg} & = & \frac{1}{4}\ \left(\sqrt{2}
G_F\right)^{1/2}~\frac{\beta(\alpha_s)}{1+\gamma_m (\alpha_s)}\
G^a_{\mu\nu} G^{a}_{\mu\nu} {\cal H}^0 \non \\
{\cal L}_{A^0 gg} & = & \frac{\alpha_s}{8\pi} \left(\sqrt{2}
G_F \right)^{1/2} G^{a}_{\mu\nu} \widetilde{G}^a_{\mu\nu} A^0
\end{eqnarray}
with $\beta$ and $\gamma_m$ defined previously. They take account of
the local interactions of the particles involved and serve as
kernels for the standard gluon and light--quark corrections.

\subsection{Higgs Production in $pp$ Collisions}
The production of \SUSY~Higgs particles at hadron colliders has received much
attention in recent years after the pioneering investigations in
Refs.~\cite{P19}. The situation is
critical since the first analyses could not ensure that the entire \MSSM~Higgs
parameter space could be covered at the LHC. Yet, high statistics analyses
appear to solve this problem if the decays to \SM~particles are dominant
\cite{P14A}.  A second similarly severe problem
has arisen from the difficulty to detect the heavy Higgs particles for masses
above a few hundred \GeV~and moderate values of $\tb$ where the production
rates in the experimentally clear
$\tau^+\tau^-$ channel are too small to be exploited in practice. However,
no final picture has emerged yet, since the detailed conclusions depend
strongly on the detector design. Additional $h^0$ decay and
production channels, based on the tagging of heavy quarks, may also
help close the hole in the parameter space \cite{A5}. \s

The dominant production process for \SUSY~Higgs particles at the LHC
is the gluon fusion mechanism.
Besides the virtual corrections, the bremsstrahlung of additional
gluons, the inelastic Compton process and quark--antiquark annihilation,
\begin{displaymath}
gg \to h^0/ H^0 / A^0 (g)\hspace{0.5cm} \mbox{~and~} \hspace{0.5cm}
gq \to h^0/ H^0 / A^0 q,~~~q\bar q \to h^0/ H^0 / A^0 g
\end{displaymath}
contribute to the Higgs production. The diagrams relevant to the various
subprocesses are the same as for the Standard Model in Fig.~1b. The parton
cross sections may thus be written
\begin{equation}
\hat\sigma_{ij} = \sigma_0 \left\{
\delta_{ig}\delta_{jg}\left[ 1+C(\tau_Q)\frac{\alpha_s}{\pi} \right]
\delta(1- \hat{\tau} ) + D_{ij}(\hat{\tau}
,\tau_Q) \frac{\alpha_s}{\pi} \Theta
(1-\hat{\tau}) \right\}
\end{equation}
for $i,j=g,q,\bar q$ and $\hat{\tau}=m_\Phi^2/\hat s$.
                                           The final result for the $pp$ cross
sections can be cast into the compact form [$\Phi = h^0, H^0, A^0$]
\begin{equation}
\sigma(pp \rightarrow h^0/H^0/A^0 +X) = \sigma_{0} \left[ 1+ C
\frac{\alpha_{s}}{\pi} \right] \tau_{\Phi} \frac{d{\cal L}^{gg}}{d\tau_{\Phi}}
+ \Delta \sigma_{gg} + \Delta \sigma_{gq} + \Delta \sigma_{q\bar{q}}
\end{equation}
with
\begin{eqnarray}
\sigma_0^{h, H} & = & \frac{G_F \alpha_s^2}{288\sqrt{2}\pi}
\left| \frac{3}{4} \sum_Q g_Q^{\cal H} A^{\cal H}_Q \right|^2 \non \\
\sigma_0^A & = & \frac{G_F \alpha_s^2}{128\sqrt{2}\pi} \left| \sum_Q g_Q^A
A^A_Q \right|^2
\end{eqnarray}
after folding the parton cross sections with the $\overline{MS}$
                                                         renormalized
quark and gluon densities [$\tau_Q = m_\Phi^2/4m_Q^2$ and $\tau_\Phi =
m_\Phi^2 / s$]. The virtual/IR and hard corrections have the same
generic form as before, eqs.~(\ref{eq:Cvirt}--\ref{eq:APKernel}).
As a result of the factorization
theorem, the parity and the specific couplings of the Higgs bosons
are not relevant for the
infrared/collinear form of the cross sections, related to interactions at
large distances.
The specific properties of the Higgs bosons affect only the non--singular
coefficients $c$ and $d$ in eqs.(\ref{eq:Cvirt}--\ref{eq:APKernel}).  \s

In the limit of large quark--loop masses compared with the Higgs masses, only
the coefficients $c$ depend on the parity of the Higgs particle,
\begin{eqnarray}
\tau_Q = m_\Phi^2/4m_Q^2 \to 0\,: \hspace{0.5cm}
c^{h^0/H^0} & \to & \frac{11}{2} \non \\
c^{A^0}     & \to & 6
\end{eqnarray}
The coefficients $d$ are universal. The next--to--leading term in the expansion
for the scalar Higgs bosons has also been calculated analytically \cite{P41}.
The form of the cross sections for the parton
subprocesses in the heavy quark--loop limit if the final states
are analyzed, is given by the same expressions as eq.~(43).
In the opposite limit of small quark--loop masses, chiral
symmetry is restored for the leading and subleading logarithmic
contributions to the coefficients, which are given by the same expressions
as eq.~(44).  \s

The final results of the $pp$ cross sections are predicted
in the subsequent figures for the LHC energy $\sqrt{s} = 14~\TeV$.
 [A brief summary is appended for $\sqrt{s} = 10$ \TeV.]
Again, the two representative values $\tb =$ 1.5 and 30
are chosen and the top mass is fixed to $m_t = 174~\GeV$. If
not stated otherwise, we have adopted the GRV
                 parameterizations of the quark
and gluon densities. For the QCD coupling we have chosen the average
value $\alpha_s^{(5)} (m_Z) = 0.117$ in the final cross sections
while the discussion of the $K$ factors is carried out consistently in
the GRV frame. [The GRV  NLO coupling is close to the lower
1$\sigma$ boundary of the global $\alpha_s$ fit.] \s

The $K$ factors, $K_{tot} = \sigma_{HO} / \sigma_{LO}$, are
defined by the ratios of the HO cross sections
to the LO cross sections. They
are shown for LHC energies in Fig.~22. They vary
little with the masses of the scalar and
pseudoscalar Higgs bosons in general, yet they depend strongly on
$\tg\beta$ as shown in Fig.~23. For small $\tg\beta$, their size is about
the same
as in the $\SM$, varying between 1.5 and 1.7; for large
$\tg\beta$ however they are in general close to unity, except
when $h^0$ approaches the $\SM$ domain. The cross sections are shown
in Fig.~24. Apart from exceptional cases, they vary in the range
between 100 and 10 pb for Higgs masses up to several
hundred \GeV. Beyond $\sim$ 300 \GeV~they drop quickly
to a level below $10^{-1}$~pb. Similarly to the
$\SM$, a factor of about 2 is lost if the $pp$
collider energies is reduced to 10 \TeV, Fig.~25. \s

The variation of the cross sections with the
renormalization/factorization scale is reduced by including
 the next--to--leading order
corrections. The dependence of the cross sections for
low masses, Fig.~26, is of order 15\%;
the $\mu$ dependence remains
monotonic. Thus the next--to--leading order corrections
stabilize the theoretical predictions for the Higgs particles in the
intermediate to large mass range, yet further improvements
must be envisaged in the future. \s

It is apparent from the previous figures that the next--to--leading order
corrections increase the production cross sections for the \SUSY~Higgs
particles, in some areas of the parameter space even strongly.

\section{Summary}
We have presented a complete next--to--leading order calculation for the
production of Higgs particles at the LHC in the Standard Model of
the electroweak interactions as well as in its minimal supersymmetric
extension. These corrections stabilize the theoretical predictions
compared with
the (ill--defined) leading--order predictions. The QCD radiative increase
the production
cross sections significantly  so that experimental opportunities
to discover and detect these fundamental particles increase.

\newpage

\nn {\bf Acknowledgements} \s

In carrying out this analysis we have benefited from discussions
with many colleagues, in particular S.~Dawson and J.~van der Bij.
The work of M.S. is supported by
Bundesministerium f\"ur Bildung und Forschung (BMBF), Bonn, Germany,
under Contract 05 6 HH 93P (5) and by EEC Program {\it Human Capital and
Mobility} through Network {\it Physics at High Energy Colliders} under
Contract CHRX--CT93--0357 (DG12 COMA).
A.D. would like to thank the DESY Theory Group for the hospitality during
the final stage of this work.

\newpage

\renewcommand{\theequation}{A.\arabic{equation}}
\setcounter{equation}{0}

\subsection*{APPENDIX A: The ${\cal H}\gamma \gamma$ and $A\gamma\gamma$
 Couplings}
\noindent In this Appendix, we summarize the complete analytical result for the
QCD corrected \CP--even ${\cal H} \gamma \gamma$
and \CP--odd $A \gamma \gamma$ vertex form--factors,
in the case of arbitrary Higgs boson and quark masses.

\vspace*{4mm}

As discussed in sections 2.1 and 3.1, the radiative QCD corrections to the
quark contribution to the two--photon Higgs boson decay amplitudes can be
written as
\begin{equation}
A_Q = A_Q^{LO} \left[1 + C_\Phi \frac{\alpha_s}{\pi} \right]
\end{equation}
where the coefficients $C_\Phi$ split into two parts,
\begin{equation}
C_\Phi = C_1^\Phi + C_2^\Phi \log \frac{\mu_Q^2}{m_Q^2}
\end{equation}
with the two functions $C_1^\Phi$ and $C_2^\Phi$ depending only on the scaling
variable $\tau$\footnote{Singularities are fixed by attributing to the
quark--loop mass a small imaginary part: $m_Q^2 \to m_Q^2 - i\epsilon$.},
\begin{eqnarray}
\tau \ \equiv \ m_\Phi^2/(4m_Q^2) \ \equiv \ \rho_\Phi/4
\end{eqnarray}
For the ${\cal CP}$--even and ${\cal CP}$--odd Higgs bosons, the coefficients
$C_1^\Phi$ and $C_2^\Phi$ are given by
\begin{eqnarray}
C_1^H & = & - \left[ 2\tau^{-1} F_0^H(\tau) \right]^{-1}~\left\{ 2\tau^{-2}
(8\tau^{-1} - 17) + 4\tau^{-2} (3\tau^{-1}-4) f(\tau)g(\tau) \right.
\nonumber \\
& & +4\tau^{-2}(\tau^{-2} - 2\tau^{-1} +3)f(\tau)
+ 24\tau^{-2} (-\tau^{-1} + 2)g(\tau) + 2\tau^{-2} (-5\tau^{-1} + 6)
l(\tau) \nonumber \\
& & + \tau^{-2} (-2\tau^{-2} + \tau^{-1} + 3)k(\tau)
+ 6\tau^{-2} (\tau^{-1} - 1)h(\tau) -8(\tau^{-2} - 3\tau^{-1} +2)I_1
\nonumber \\
& & + 4
\tau^{-1} (2\tau^{-1} - 3)[I_2-I_3+I_4] + 16\tau^{-1} (\tau^{-1} - 1)I_5
\left.  \right\} \nonumber \\
C_2^H & = & -3\tau^{-1}~\left[ F_0^H (\tau) \right]^{-1}~ \left\{
(2\tau^{-1}-1) f(\tau) -g(\tau) -1 \right\} \nonumber \\
C_1^A & = & \left[3\tau^{-1} F_0^A (\tau) \right]^{-1}~ \left\{ 24 \tau^{-2}+
16\tau^{-2} f(\tau)g(\tau) \right.  + 6\tau^{-2} (\tau^{-1} - 2)f(\tau) -
36\tau^{-2} g(\tau) \nonumber \\
& & - 12\tau^{-2} l(\tau)
- 3\tau^{-2} (\tau^{-1} + 1) k(\tau) + 6\tau^{-2} h(\tau) +8(-\tau^{-1} +
2)I_1 \nonumber \\
& & + 12\tau^{-1} [I_2-I_3+I_4] + 16\tau^{-1} I_5 \left.  \right\} \nonumber
\\
C_2^A & = & 2\tau^{-1} \left[F_0^A(\tau)\right]^{-1}~ \left\{ f(\tau)
-g(\tau)/ (\tau^{-1}-1) \right\}
\end{eqnarray}
with
\begin{eqnarray}
F_0^H (\tau) = \frac{3}{2} \tau^{-1} \left[ 1 + (1-\tau^{-1}) f(\tau) \right]
\ \ \  \mbox{\rm and} \ \ \ F_0^A (\tau) = \tau^{-1} f(\tau)
\end{eqnarray}
In terms of the auxiliary variables
\begin{eqnarray}
\alpha_\pm = (1 \pm \sqrt{1-\tau^{-1}} )/2
\end{eqnarray}
the functions $f,g, l,k$ and $h$ are defined by
\begin{eqnarray}
f(\tau)& =& \left\{
\begin{array}{ll}  \displaystyle
\arcsin^2\sqrt{\tau} & \tau\leq 1 \\
\displaystyle -\frac{1}{4}\left[ \log\frac{\alpha_+}
{\alpha_-}-i\pi \right]^2 \hspace{0.5cm} & \tau>1
\end{array} \right. \nonumber \\ \nonumber \\
g(\tau)& =& \left\{
\begin{array}{ll}  \displaystyle
\sqrt{\tau^{-1}-1} \arcsin \sqrt{\tau} & \tau\leq 1 \\
\displaystyle \frac{\sqrt{1-\tau^{-1}}}{2}\left[ \log\frac{\alpha_+}
{\alpha_-}-i\pi \right] \hspace{0.5cm} & \tau>1
\end{array} \right. \nonumber \\ \nonumber \\
l(\tau) & = & Li_3(1/\alpha_+) + Li_3(1/\alpha_-) \nonumber \\ \nonumber \\
k(\tau) & = & S_{1,2}(1/\alpha_+) + S_{1,2}(1/\alpha_-)
\nonumber \\ \nonumber \\
h(\tau) & = & 4\left[ S_{1,2} \left( \frac{\alpha_+}{\alpha_-} \right) +
S_{1,2} \left( \frac{\alpha_-}{\alpha_+} \right) \right] + 2\left[ Li_3 \left(
\frac{\alpha_+}{\alpha_-} \right) + Li_3 \left( \frac{\alpha_-}{\alpha_+}
\right)  \right] + 2 \zeta(3)
\end{eqnarray}
Here, $Li_2 , Li_3$ and $S_{1,2}$ are polylogarithms, defined \cite{P54} as
\begin{eqnarray}
Li_2 (x) & = & \int_0^1 \frac{dy}{y} \log(1-xy)  \non \\
Li_3 (x) & = & \int_0^1 \frac{dy}{y} \log y \log(1-xy)  \non \\
S_{1,2} (x) & = & \frac{1}{2} \int_0^1 \frac{dy}{y} \log^2(1-xy)
\end{eqnarray}
The expressions of $I_{1,\ldots,5}$,
which have
been reduced from four-- and five--dimensional down to one--dimensional
Feynman integrals,
are much more involved:
%
%
\begin{eqnarray}
I_1 & = &
\int_0^1 \frac{dx}{\rho} \frac{\log[1-\rho x(1-x)]}{1-\rho
x(1-x)} \left\{ Li_2(\rho x) -Li_2[\rho x(1-x)] + Li_2\left(\frac{-\rho
x^2}{1-\rho x} \right) \right.  \nonumber \\
& & \hspace{4cm} \left.  - Li_2\left(\frac{-\rho x}{1-\rho x} \right) -
log[x(1-x)] \log[1-\rho x] \right. \nonumber \\
& & \left. + Li_2\left(\frac{\rho x^2}{1-\rho x(1-x)} \right) -
Li_2\left(\frac{-\rho x(1-x)}{1-\rho x(1-x)} \right) \right.  \nonumber \\
& & +Li_2\left(\frac{1-\rho x}{1-\frac{x}{\alpha_+}} \right)
+Li_2\left(\frac{1-\rho x}{1-\frac{x}{\alpha_-}} \right)
-Li_2\left(\frac{1}{1-\frac{x}{\alpha_+}} \right)
-Li_2\left(\frac{1}{1-\frac{x}{\alpha_-}} \right) \nonumber \\
& & \left.  + \frac{\log^2(1-\rho x)}{2} + \log(1-\rho x) \log\left(\frac{\rho
x^2}{1-\rho x(1-x)} \right) \right\} \nonumber \\ \nonumber \\
& + & \int_0^1 \frac{dx}{\rho [1-\rho x(1-x)]} \left\{ Li_3[\rho
x(1-x)] + Li_3\left(\frac{1-\rho x}{-\rho x} \right) - Li_3\left(\frac{1-\rho
x}{-\rho x^2} \right) \right.  \nonumber \\
& & + S_{1,2}(1-x) -S_{1,2}\left(\frac{1-x}{-x} (1-\rho x) \right)
\nonumber \\
& & \left.  - Li_2\left(\frac{1-\rho x}{-\rho x}\right) \log x +\frac{\log^2
x}{2} \log(\rho x) \right. \nonumber \\ \nonumber \\
& & \left. +
Li_3\left(\frac{-x}{1-x}\right) - Li_3\left(\frac{-x}{(1-x)(1-\rho x)} \right)
\right.  \nonumber \\
& & + S_{1,2}[\rho x(1-x)] + S_{1,2} (\rho x) -S_{1,2}\left(\frac{-\rho
x^2}{1-\rho x} \right) \nonumber \\
& & \left.  - \log(1-\rho x) Li_2\left(
\frac{-x}{1-x} \right) +\frac{\log (1-x)}{2} \log^2(1-\rho x) \right\}
\end{eqnarray}
%
%
\begin{eqnarray}
I_2 & = &
\int_0^1 \frac{dx}{\rho^2 x^2} \left\{ \rho x \left[ 6 -
\log\left( 1-\frac{x}{\alpha_+}\right)
\log\left( 1-\frac{x}{\alpha_-}\right) \right] \right. \nonumber \\
& & -\rho x(1-x) \left[ Li_2(\rho x(1-x))-2Li_2\left(\frac{x}{\alpha_+}\right)
-2Li_2\left(\frac{x}{\alpha_-}\right) \right] \nonumber \\
& & + 2\left[ 2\sqrt{\rho x(1-x)}
\left( \log\left(1-\sqrt{\rho x(1-x)}\right)
-\log\left(1+\sqrt{\rho x(1-x)}\right) \right) \right. \nonumber \\
& & \left. + \rho x (\alpha_+-\alpha_-)
\left(\log\left(1-\frac{x}{\alpha_+}\right)
-\log\left(1-\frac{x}{\alpha_-}\right) \right) \right] \nonumber \\
& & +[2-\rho x(1-x)] \left[ S_{1,2}(\sqrt{\rho x(1-x)})
+S_{1,2}(-\sqrt{\rho x(1-x)}) \right. \nonumber \\
& & +S_{1,2}\left(\frac{-\sqrt{\rho x(1-x)}}{1-\sqrt{\rho x(1-x)}} \right)
+S_{1,2}\left(\frac{\sqrt{\rho x(1-x)}}
{1+\sqrt{\rho x(1-x)}} \right)
-S_{1,2}\left(\frac{x}{\alpha_+} \right)
-S_{1,2}\left(\frac{x}{\alpha_-} \right) \nonumber \\
& & \left. \left. -S_{1,2}\left(\frac{-x}{\alpha_+-x} \right)
-S_{1,2}\left(\frac{-x}{\alpha_--x} \right) \right]\right\} \nonumber \\
\nonumber \\
& + & \int_0^1 dx \left\{ \frac{1-x}{\rho} \left[ K_9\left( -(1-x),
\frac{\rho x(1-x)}{1-\rho x(1-x)} \right) -K_6\left(
\frac{\rho x(1-x)}{1-\rho x(1-x)}, -(1-x) \right) \right. \right. \nonumber \\
& & \left. \hspace{3cm} -\log(1-x) K_{11}\left(
\frac{\rho x(1-x)}{1-\rho x(1-x)}, -(1-x) \right) \right] \nonumber \\
& & -(1-x)\frac{1+\rho x}{\rho^2 x^2}\left[ K_3\left( 1,-(1-x),1,
\frac{\rho x(1-x)}{1-\rho x(1-x)} \right) \right. \nonumber \\
& & \hspace{3cm} -K_1\left( 1,
\frac{\rho x(1-x)}{1-\rho x(1-x)},1,-(1-x)\right) \nonumber \\
& & \left. \hspace{3cm} -\log(1-x) K_2\left( 1,
\frac{\rho x(1-x)}{1-\rho x(1-x)}, 1,-(1-x) \right) \right] \nonumber \\
& & +\frac{1-x}{\rho^2 x^2 (1-\rho x)} \frac{1-\rho x(1-2x)}{1-\rho x(1-x)}
\left[ K_4\left(
\frac{\rho x(1-x)}{1-\rho x(1-x)}, -(1-x),
-\frac{(1-x)(1-\rho x)}{1-\rho x(1-x)} \right) \right. \nonumber \\
& & -K_1\left( 1,
\frac{\rho x(1-x)}{1-\rho x(1-x)}, 1,-(1-x),
-\frac{(1-x)(1-\rho x)}{1-\rho x(1-x)} \right) \nonumber \\
& & \left. -\log(1-x) K_2\left( 1,
\frac{\rho x(1-x)}{1-\rho x(1-x)}, 1,
-\frac{(1-x)(1-\rho x)}{1-\rho x(1-x)} \right) \right] \nonumber \\
& & -2\frac{1-x}{\rho x(1-\rho x)} \left[ K_{10} \left(
\frac{\rho x(1-x)}{1-\rho x(1-x)}, -(1-x) \right) - K_7\left( 1,
\frac{\rho x(1-x)}{1-\rho x(1-x)}\right) \right. \nonumber \\
& & \left. \left. + \log(1-x) \left( 1+\frac{\log [
1-\rho x(1-x)]} {\rho x(1-x)} \right) \right] \right\} \nonumber \\
& - & 2 \int_0^1 \frac{dx (1-x)}{\rho x} \left\{ \log[1-\rho x(1-x)]
\left[ \log\left(\rho (1-x)^2\right) - \frac{1}{2} \log [1-\rho x(1-x)]
\right] \right. \nonumber \\
& & + \log\left[ 1+\sqrt{\rho x(1-x)}\right]
\log\left[ 1-\sqrt{\rho x(1-x)}\right] + Li_2(\rho x(1-x)) + Li_2\left(
1-\frac{1-x}{\alpha_+} \right) \nonumber \\
& & \left. + Li_2\left( 1-\frac{1-x}{\alpha_- } \right)
- Li_2\left( \frac{1}{1-\frac{1-x}{\alpha_+ }} \right)
- Li_2\left( \frac{1}{1-\frac{1-x}{\alpha_- }} \right) \right\}
\end{eqnarray}
%
%
\begin{eqnarray}
I_3 &= & \int_0^1 dx \left\{ \frac{\log[1-\rho x(1-x)]}{\rho x} \right.
+ \frac{x(1-2x)}{2[1-\rho x(1-x)]} \log^2[1-\rho x(1-x)] \nonumber \\
& & + \frac{2-\rho x -\rho^2 x^2(1-x)(1-2x)}{\rho^2 x(1-x) [1-\rho x(1-x)]}
\log\left[1+\sqrt{\rho x(1-x)}\right] \log\left[1-\sqrt{\rho x(1-x)}\right]
\nonumber \\
& & + \frac{2}{\rho \sqrt{\rho x(1-x)}} \left[\log\left(1+\sqrt{\rho
x(1-x)}\right) - \log\left(1-\sqrt{\rho x(1-x)}\right)\right] \nonumber \\
& & + \frac{(2-\rho x)(1- \rho x)}{\rho^2 x^2 [1-\rho x(1-x)]}
\log\left(1-\frac{x}{\alpha_+}\right) \log\left( 1-\frac{x}{\alpha_-} \right)
 \nonumber \\
& & + \frac{\alpha_+-\alpha_-}{\rho x} \left[ \log\left(
1-\frac{x}{\alpha_+}\right) - \log\left( 1-\frac{x}{\alpha_-} \right) \right]
-\frac{3x(1-2x)}{2[1-\rho x(1-x)]} Li_2[\rho x(1-x)]  \nonumber \\
& &+\frac{x(1-2x)}{1-\rho x(1-x)]}\left[ Li_2\left(\frac{x}{\alpha_+} \right) +
Li_2 \left( \frac{x}{\alpha_-} \right) -\log[1-\rho x(1-x)] \log[\rho (1-x)^2]
\right. \nonumber \\
& & + Li_2\left( \frac{-\alpha_+}{\alpha_- - x}\right) + Li_2\left( \frac{-
\alpha_-}{\alpha_+ - x}\right) \left.  \left.  - Li_2\left( \frac{\alpha_-
-x}{-\alpha_+}\right) - Li_2\left( \frac{\alpha_+ -x}{-\alpha_-}\right)
\right] \right\}
\end{eqnarray}
%
%
\begin{eqnarray}
I_4 &= & \int_0^1 dx \left\{ -\frac{1}{\rho x} \left[ Li_3\left( \frac{-\rho
x(1-x)}{1-\rho x(1-x)} \right) + S_{1,2}\left( \frac{-\rho x(1-x)}{1-\rho
x(1-x)} \right) \right] \right. +(1-2x) \log x  \nonumber \\
& & \times \left[ \frac{\log\left(1+\sqrt{\rho
x(1-x)}\right)\log\left(1-\sqrt{\rho x(1-x)}\right)}{\rho x(1-x)}
\left.  - \frac{Li_2[\rho x(1-x)]}{2[1-\rho x(1-x)]} \right] \right\}
\end{eqnarray}
%
%
\begin{eqnarray}
I_5 = \int_0^1 \frac{dx}{1-\rho x} \left\{ \alpha_+
\log\left(1-\frac{x}{\alpha_+} \right) + \alpha_-
\log\left(1-\frac{x}{\alpha_-} \right) \right\} \log\left( \frac{1-\rho
x(1-x)}{x} \right)
\end{eqnarray}

\vspace*{0.3mm}

In terms of the variables $\alpha = (ad-bc)/d$ and $\beta = b/d$, and the
functions
\begin{eqnarray}
F_1(a,b) & = & S_{1,2} (-a) + S_{1,2}(-b) -S_{1,2}\left( \frac{b-a}{1+b}
\right) -Li_3\left( \frac{b}{a} \frac{1+a}{1+b} \right) +Li_3\left(
\frac{b}{a} \right) \nonumber \\
& & +\log\left(\frac{1+a}{1+b}\right) Li_2\left( \frac{b}{a}\frac{1+a}{1+b}
\right) +\frac{1}{2} \log^2\left( \frac{1+a}{1+b}\right) \log\left(
\frac{a-b}{a(1+b)} \right) \\
F_2(a,b,c,d) & = & \frac{1}{d}\left\{ \log^2\alpha
\log\left(\frac{c+d}{c}\right) +2\log\alpha\left[ Li_2 \left( -
\frac{\beta c}{\alpha}\right) -Li_2\left( -\frac{\beta}{\alpha} (c+d) \right)
\right] \right. \nonumber \\
& & \left. +2S_{1,2}\left(-\frac{\beta}{\alpha}(c+d)\right)-2S_{1,2} \left(-
\frac{\beta c}{\alpha} \right) \right\}
\end{eqnarray}
the expressions $K_i$, which appear in the integral $I_2$, are given by
[$K_5$ and $K_8$ will be used later on]
\begin{eqnarray}
K_1(a,b,c,d) & = & \frac{1}{d}\left\{ \log\alpha \log\left(-\frac{c}{d}\right)
\log\left(\frac{c+d}{c}\right) + \log\alpha\left[ \zeta(2) - Li_2 \left(
\frac{c+d}{c}\right) \right] \right. \nonumber \\
& & +\log\left(-\frac{c}{d}\right) \left[ Li_2\left( -
\frac{\beta c}{\alpha}\right) -Li_2\left( -\frac{\beta}{\alpha} (c+d) \right)
\right] \nonumber \\
& & \left. -Li_3\left(-\frac{\beta c}{\alpha} \right) -
S_{1,2} \left(-\frac{\beta c}{\alpha} \right) + F_1\left[ -\frac{c+d}{c},
\frac{\beta}{\alpha} (c+d) \right] \right\} \nonumber \\
K_2(a,b,c,d) & = & \frac{1}{d}\left\{ \log\alpha
\log\left(\frac{c+d}{c}\right) + Li_2 \left( -
\frac{\beta c}{\alpha}\right) -Li_2\left( -\frac{\beta}{\alpha} (c+d) \right)
\right\} \nonumber \\
K_3(a,b,c,d) & = & \frac{1}{2b} \left\{ \log^2(a+b) \log(c+d) - \log^2 a \log
c -d F_2(a,b,c,d) \right\} \nonumber \\
K_4(a,b,c) & = & \frac{1}{c} \left\{ \log\left( \frac{c-a}{c} \right)
\log\left( \frac{c-b}{c} \right) \log(1+c) \right. \nonumber \\
& & + \log\left( \frac{c-a}{c} \right) \left[ Li_2\left( \frac{-b}{c-b}
\right) - Li_2\left( -b \frac{1+c}{c-b} \right) \right]
+ F_1\left( a\frac{1+c}{c-a}, b\frac{1+c}{c-b} \right) \nonumber \\
& & + \log\left( \frac{c-b}{c} \right) \left[ Li_2\left( \frac{-a}{c-a}
\right) - Li_2\left( -a \frac{1+c}{c-a} \right) \right]
- \left. F_1\left( \frac{a}{c-a}, \frac{b}{c-b} \right) \right\} \nonumber \\
K_5(a) & = & a\left\{1-Li_2(-a) \right\} -(1+a) \log(1+a) \nonumber \\
K_6(a,b) & = & \frac{1}{a-b} \left\{ Li_2(-a) -\frac{a}{b} Li_2(-b) \right\} -
K_2(1,a,1,b) \nonumber \\
K_7(a,b) & = & 2+\frac{a}{b} \left\{ \log a + Li_2 \left( -\frac{b}{a} \right)
\right\} - \frac{a+b}{b} \log(a+b) \nonumber \\
K_8(a,b) & = & -\log(1+a)\log(1+b) -a Li_2(-b) \nonumber \\
& & -a^2 K_2(1,b,1,a) - b^2 K_2(1,a,1,b) \nonumber \\
K_9(a,b) & = & \frac{1}{a} \left\{ \frac{b}{a-b} \left[ \log \left(
\frac{1+a}{1+b} \right) + \frac{1}{2} \log^2(1+a) -b K_2(1,a,1,b) \right]
\right. \nonumber \\
& & \left. -\log(1+b) \frac{1+\log(1+a)}{1+a} \right\} \nonumber \\
K_{10}(a,b) & = & 2 + \frac{1+a}{a} \log(1+a) \left\{ \log(1+b) -1 \right\} -
\frac{1+b}{b} \log(1+b) \nonumber \\
& & + \frac{a-b}{ab} \left\{ \log\left( \frac{b-a}{b} \right) \log(1+b) +
Li_2\left( \frac{-a}{b-a} \right) - Li_2 \left( -a \frac{1+b}{b-a} \right)
\right\} \nonumber \\
K_{11} (a,b) & = & \frac{1}{b-a} \left\{ \frac{a}{b} \log(1+b) -
\frac{1+a}{1+b} \log(1+a) \right\}
\end{eqnarray}

\renewcommand{\theequation}{B.\arabic{equation}}
\setcounter{equation}{0}

\subsection*{APPENDIX B: The Infrared Regularized Virtual Corrections for the
 ${\cal H}gg$ and $Agg$ Couplings}

\noindent The complete analytical expressions for the virtual, infrared
regularized, QCD radiative corrections to the ${\cal H}gg$ and $Agg$ couplings
are summarized in this appendix. As
discussed in section 2.3 and 3.3, the virtual corrections split into an
infrared part $\pi^2$, a logarithmic part depending on the renormalization
scale $\mu$ and a finite piece depending on $\tau$
\begin{eqnarray}
C = \pi^2 +   \frac{33-2N_F}{6}~\log \frac{\mu^2}{m_\Phi^2} + c_\Phi
(\tau)
\end{eqnarray}
Again the coefficient $c_\Phi$ can be split into two parts
\begin{eqnarray}
c_\Phi = \Re e \left\{ \frac{ \sum_Q F_0^\Phi (\tau) \left( B_1^\Phi
+B_2^\Phi \log\frac{\mu_Q^2}{m_Q^2} \right)}{\sum_Q F_0^\Phi (\tau)} \right\}
\end{eqnarray}
where $\mu_Q$ is the scale at which the quark mass is defined; the coefficients
$B_1, B_2$ read in the case of ${\cal CP}$--even and ${\cal CP}$--odd Higgs
bosons,
\begin{eqnarray}
B_1^H & = & -\left[16\tau^{-1} F_0^H(\tau) \right]^{-1}~\left\{ 32\tau^{-2}
(-3\tau^{-1}+1) f(\tau)g(\tau) \right. \nonumber \\
& & + 40\tau^{-2} (-2\tau^{-2} + 4\tau^{-1} +
3)f(\tau)
+ 192\tau^{-2} (4\tau^{-1} - 5) g(\tau) \nonumber \\
& & + 216\tau^{-2} (-2\tau^{-1} +
3)r(\tau) + 36\tau^{-2} (-2\tau^{-1} + 3)p(\tau) \nonumber \\
& & + 4\tau^{-2} (18\tau^{-2} - 49\tau^{-1} + 30) l(\tau) + 2\tau^{-2}
(38\tau^{-2}-199\tau^{-1} + 186)k(\tau) \nonumber \\
& & + 3\tau^{-2} (-18\tau^{-2} + 65\tau^{-1} - 53)h(\tau) + 144(\tau^{-1}-1)[2
I_6 - \tau I_8] \nonumber \\
& & + 16\tau^{-2} (-29\tau^{-1} + 20)
+128(-\tau^{-2} + 3\tau^{-1} - 2)I_1 \nonumber \\
& & - 8(2\tau^{-1} - 3)[\tau^{-1}
I_2-10\tau^{-1} I_3+\tau^{-1} I_4 + 18 I_7] - 32\tau^{-1} (\tau^{-1} - 1)I_5
\left.  \right\} \nonumber \\
B_2^H & = & -6\tau^{-1}~\left[ F_0^H(\tau) \right]^{-1} \left\{(2\tau^{-1}-1)
f(\tau) -g(\tau) -1 \right\} \nonumber \\
B_1^A & = & \left[24\tau^{-1} F_0^A(\tau) \right]^{-1} \{ -32\tau^{-2} f(\tau)
g(\tau) -120\tau^{-2} (\tau^{-1} + 1)f(\tau) +1152 \tau^{-2} g(\tau)
\nonumber \\
& & - 648\tau^{-2} r(\tau)
- 108 \tau^{-2} p(\tau) +12\tau^{-2} (9\tau^{-1} - 10)l(\tau) + 6\tau^{-2}
(19\tau^{-1} - 62) k(\tau) \nonumber \\
& & + 3\tau^{-2} (53-27\tau^{-1})h(\tau)
+128(2-\tau^{-1})I_1 -24[\tau^{-1} I_2-10\tau^{-1} I_3+\tau^{-1} I_4 + 18
I_7] \nonumber \\
& & - 32\tau^{-1} I_5 + 144[2 I_6 - \tau^{-1} I_8] - 696\tau^{-2} \left.
\right\} \nonumber \\
B_2^A & = & 4\tau^{-1} \left[F_0^A(\tau)\right]^{-1}~\left\{ f(\tau)
-g(\tau)/(\tau^{-1}-1) \right\}
\end{eqnarray}
The functions $F_0^H, F_0^A$ are given in eq.~(A.5), while the functions
$f,g,l,k$ and $h$ are given in eq.~(A.7); the two remaining functions
$r$ and $p$ read
\begin{eqnarray}
r(\tau) & = & -\frac{1}{2}\sqrt{1-\tau^{-1}} \left[ Li_2\left(
\frac{\alpha_+}{\alpha_+-\alpha_-} \right) - Li_2\left(
\frac{\alpha_-}{\alpha_--\alpha_+} \right) \right]+ \frac{g(\tau)}{2} \log
\left[ 4 (1-\tau) \right] \nonumber \\ \nonumber \\
p(\tau) & = & \sqrt{1-\tau^{-1}} \left[ Li_2(1/\alpha_+) - Li_2(1/\alpha_-)
\right]
\end{eqnarray}

\vspace*{2mm}

The integrals $I_1$ to $I_5$ were presented in eqs.~(A.9--A.13); the integrals
$I_6 , I_7$ and
$I_8$ are given by
\begin{eqnarray}
I_6 & = & \frac{2}{\rho} \int_0^1 \frac{dx}{\rho x(1-x)} \left\{
\frac{8}{3} \log^3[1-\rho x(1-x)] - 2\log(1-x) \log^2[1-\rho x(1-x)] \right.
\nonumber \\
& & +3\log[1-\rho x(1-x)] Li_2(\rho x(1-x))+8S_{1,2}(\rho x(1-x)) \nonumber \\
& & +2\frac{\rho x^2}{1-\rho x} \left[ K_4\left( -\rho x(1-x), -x,
\frac{\rho x^2}{1-\rho x} \right) -K_{17}\left( 1,-\rho x(1-x),1,\frac{\rho
x^2}{1-\rho x} \right) \right] \nonumber \\
& & \left. +4 \rho x(1-x) K_3\left[1,-\rho x(1-x),1-\rho x,\rho x^2\right]
\right\}
\end{eqnarray}
\begin{eqnarray}
I_7 &=& \frac{2}{\rho} \int_0^1 \frac{dx}{\rho} \left\{
\frac{8}{3} \log^3[1-\rho x(1-x)] - 2\log(1-x) \log^2[1-\rho x(1-x)] \right.
\nonumber \\
& & +3\log[1-\rho x(1-x)] Li_2(\rho x(1-x))+8S_{1,2}(\rho x(1-x)) \nonumber \\
& & +2\frac{\rho x^2}{1-\rho x} \left[ K_4\left( -\rho x(1-x), -x,
\frac{\rho x^2}{1-\rho x} \right) -K_{17}\left( 1,-\rho x(1-x),1,\frac{\rho
x^2}{1-\rho x} \right) \right] \nonumber \\
& & \left. +4 \rho x(1-x) K_3\left[1,-\rho x(1-x),1-\rho x,\rho x^2\right]
\right\} \nonumber \\ \nonumber \\
&+& \frac{2}{\rho} \int_0^1 dx~x(1-x) \left\{ \frac{2}{\rho x(1-x)} \left[
\log[1-\rho x(1-x)] Li_2\left( \frac{-\rho x(1-x)}{1-\rho x(1-x)} \right)
\right. \right. \nonumber \\
& & \left. \hspace{5cm} - S_{1,2} \left(\frac{-\rho x(1-x)}{1-\rho x(1-x)}
\right) \right] \nonumber \\
& & +F_2\left[ 1,-x,1,-\rho x(1-x)\right] +F_2\left[ 1,-\rho x(1-x), 1-\rho
x,\rho x^2 \right] \nonumber \\
& & -F_2\left[ 1,-x,1-\rho x,\rho x^2 \right] + K_1\left[ 1,-x,1,-\rho x(1-x)
\right] \nonumber \\
& & - K_1\left[ 1,-\rho x(1-x),1-\rho x,\rho x^2 \right] - K_1 \left[
1,-x,1-\rho x,\rho x^2\right] \nonumber \\
& & -2 K_3[1,-\rho x(1-x),1,-x] - K_{17} \left[ 1,-x,1,-\rho
x(1-x)\right] \nonumber \\
& & + K_{17}\left[ 1,-x,1-\rho x,\rho x^2\right] - K_{17}\left[ 1,-\rho
x(1-x),1-\rho x, \rho x^2 \right] \nonumber \\
& & \left. - K_{18} \left[ 1,-\rho x(1-x) \right] + K_{18} \left[ 1-\rho x,
\rho x^2 \right] \right\} \nonumber \\ \nonumber \\
&+& \frac{2}{\rho}
\int_0^1 dx~x \left\{ \log[1-\rho x(1-x)] \left[ -\frac{2}{\rho
x} - \frac{2}{1-\rho x} \right] \right. \nonumber \\
& & + \log^2[1-\rho x(1-x)] \left[ \frac{3}{2\rho
x} + \frac{1}{\rho(1-x)} + \frac{3}{2(1-\rho x)}\right] \nonumber \\
& & +\log\left( \frac{1-\rho x(1-x)}{1-\rho x} \right) \left[
-2\frac{2-x}{\rho x^2} + 2\frac{1-x}{x} \right] + \frac{1-x}{2x} \frac{1-\rho
x}{\rho x} \log^2\left( \frac{1-\rho x(1-x)}{1-\rho x} \right) \nonumber \\
& & +\frac{2}{\rho(1-x)} Li_2 (\rho x(1-x)) + 2 \frac{1-x}{\rho x^2}
Li_2\left( \frac{\rho x^2}{1-\rho x(1-x)}\right) + 2 \frac{\rho x}{1-\rho x}
\log x \nonumber \\
& & +\frac{1}{2} \frac{\rho x}{1-\rho x} \log^2 x - F_2 \left[ 1,-\rho x(1-x),
1-\rho x, \rho x^2\right] + F_2\left[ 1,-x,1-\rho x, \rho x^2\right] \nonumber
\\
& & + K_1 [1,-x,1-\rho x, \rho x^2] + K_1[1,-\rho x(1-x),1-\rho x,\rho x^2]
\nonumber \\
& & +2(1-x) K_2[1,-\rho x(1-x), 1-\rho x, \rho x^2] + 2K_2[1,-x,1,-\rho
x(1-x)] \nonumber \\
& & + 2\rho x(1-x) K_2[1,-x,1-\rho x,\rho x^2] \nonumber \\
& & +2\frac{\rho x(1-x)}{1-\rho x} \left[ K_2[1,-(1-x),1,-\rho x(1-x)] +
K_2[1,-\rho x(1-x),1,-(1-x)] \right] \nonumber \\
& & + 2(1-x) K_2[1,-\rho x(1-x),1,-x] \nonumber \\
& & -\frac{1-\rho x(1-x)}{(1-\rho x)^2} \left[ K_6\left( -x,\frac{\rho
x^2}{1-\rho x} \right) + K_6\left( -\rho x(1-x),\frac{\rho x^2}{1-\rho
x}\right) \right] \nonumber \\
& & -2 \frac{1-\rho x(1-x)}{(1-\rho x)^2} \left[ K_{11}\left( -x,\frac{\rho
x^2}{1-\rho x} \right) + K_{11} \left( -\rho x(1-x), \frac{\rho x^2}{1-\rho x}
\right) \right] \nonumber \\
& & -\frac{1-\rho x(1-x)}{(1-\rho x)^2} \left[ K_{12}\left( -x,\frac{\rho
x^2}{1-\rho x}\right) - K_{12}\left( -\rho x(1-x), \frac{\rho x^2}{1-\rho x}
\right) \right] \nonumber \\
& & - K_{17}[1,-x,1-\rho x,\rho x^2] + K_{17}[1,-\rho x(1-x),1-\rho x,\rho
x^2] \nonumber \\
& & - K_{18}[1-\rho x,\rho x^2] + [1-\rho x(1-x)] K_{19} [1-\rho x, \rho x^2]
\nonumber \\
& & \left. -[1-\rho x(1-x)] \left[ K_{20}[ 1,-\rho x(1-x), 1-\rho x, \rho x^2]
- K_{20}[1,-x,1-\rho x,\rho x^2] \right] \right\}
\end{eqnarray}
\begin{eqnarray}
I_8 & = & \frac{2}{\rho} \int_0^1 \frac{dx}{x} \left\{ -\frac{1}{2} \log(1-x)
\log^2[1-\rho x(1-x)] + 3S_{1,2}(\rho x(1-x)) \right. \nonumber \\
& & -2 \log[1-\rho x(1-x)] Li_2 \left( \frac{-\rho x(1-x)}{1-\rho x(1-x)}
\right) \nonumber \\
& & + F_1\left[ -\rho x(1-x), \frac{1-x}{x} (1-\rho x) \right]
-F_1\left[ -\rho x(1-x), \frac{1-x}{x} \right] \nonumber \\
& & \left. + \rho x(1-x)
K_3\left[ 1,-\rho x(1-x),x,(1-x)(1-\rho x)\right] \right\}
\end{eqnarray}
While the functions $K_1$--$K_{11}$ are defined in eqs.(A.16), the remaining
functions follow from
\begin{eqnarray}
K_{12}(a,b) & = & \frac{1}{a-b} \left\{ \frac{1+a}{1+b} \log^2(1+a) - 2a
K_2(1,a,1,b) \right\} \nonumber \\
K_{13}(a,b,c) & = & -\frac{\log(1+a)\log(1+b)}{c(1+c)} +\frac{1}{c} \left\{
\frac{a^2}{a-c} K_2(1,b,1,a) \right. \nonumber \\
& & \left. -\frac{ac}{a-c} K_2(1,b,1,c)
+\frac{b^2}{b-c} K_2(1,a,1,b) -\frac{bc}{b-c} K_2(1,a,1,c) \right\} \nonumber
\\
K_{14}(a) & = & a - (1+a) \log(1+a) \nonumber \\
K_{15}(a) & = & -(1+a) \log^2(1+a) -2a Li_2(-a) \nonumber \\
K_{16}(a,b) & = & K_7(a+b,-b) \nonumber \\
K_{17}(a,b,c,d) & = & K_1(a+b,-b,c+d,-d) \nonumber \\
K_{18}(a,b) & = & \frac{1}{b} \left\{ \log\left( \frac{a+b}{a} \right) \left[
Li_2\left( \frac{-b}{a} \right) -\zeta(2) \right] + S_{1,2} \left(
\frac{-b}{a} \right) -2 Li_3 \left( \frac{-b}{a} \right) \right\} \nonumber \\
K_{19} (a,b) & = & \frac{1}{ab} Li_2\left( \frac{b}{a+b} \right) -
\frac{1}{b(a+b)} Li_2\left( -\frac{b}{a} \right) -\frac{\zeta (2)}{a(a+b)}
\nonumber \\
K_{20}(a,b,c,d) & = & \frac{1}{d} \left\{ \frac{b}{\alpha d} \left[ Li_2
\left( \frac{b}{a+b} \right) - Li_2 \left( \frac{d}{c+d} \right) \right]
\right. \nonumber \\
& & \left. +\frac{1}{c+d} Li_2\left( \frac{b}{a+b} \right) - \frac{d}{c+d}
K_2(a,b,c,d) \right\}
\end{eqnarray}

\renewcommand{\theequation}{C.\arabic{equation}}
\setcounter{equation}{0}

\subsection*{APPENDIX C: The Real Corrections for pp~$\to {\cal H},A$ and
 ${\cal H},A \to$~gg}

\noindent Finally, we give here the complete analytical expressions for the
real corrections to the processes $pp \to {\cal H}/A$ and ${\cal H}/A \to
gg$. We
start with the corrections to the production process and define the variables
\begin{eqnarray}
\rho = \frac{m_H^2}{m_Q^2} \ , \hspace{1cm}
S    = \frac{\hat s}{m_Q^2} \ , \hspace{1cm}
T    = \frac{\hat t}{m_Q^2} \ , \hspace{1cm}
U    = \frac{\hat u}{m_Q^2}
\end{eqnarray}
\begin{eqnarray}
\hat s = \frac{m_H^2}{\hat\tau} \ , \hspace{1cm} \hat t =
-\hat s(1-\hat\tau)v \ , \hspace{1cm} \hat u = -\hat s(1-\hat\tau)(1-v)
\end{eqnarray}
\begin{eqnarray}
\tau_Q = \frac{\rho}{4} \ , \hspace{1cm}
\tau_s = \frac{S}{4} \ , \hspace{1cm}
\tau_t = \frac{T}{4} \ , \hspace{1cm}
\tau_u = \frac{U}{4}
\end{eqnarray}
The coefficients $d_{q \bar{q}}, d_{gq}$ and $d_{gg}$ which appear in
the real QCD corrections, eq.~(40), for the Higgs production, can be cast into
the form
\begin{eqnarray}
d_{q\bar q}(\hat\tau,\tau_Q) & = &
\frac{2}{3\left|\sum_Q F_0^\Phi(\tau_Q)\right|^2} (1-\hat\tau)^3 \left| \sum_Q
{\cal A}^\Phi_{qqg} (S) \right|^2 \nonumber \\ \nonumber \\
d_{gq}(\hat\tau,\tau_Q) & = & \frac{2}{3}\hat\tau^2 +
\frac{2}{3} \hat\tau^2 \int_0^1 \frac{dv}{v} \left\{ -1
-2\frac{1-\hat\tau}{\hat\tau^2} \right. \nonumber \\
& & \left. + \frac{1+(1-\hat\tau)^2(1-v)^2}{\hat\tau^2}
\left| \frac{3}{2\sum_Q F_0^\Phi(\tau_Q)} \sum_Q {\cal A}^\Phi_{qqg} (T)
\right|^2 \right\}  \nonumber \\
d_{gg}(\hat\tau,\tau_Q) & = & \frac{3}{1-\hat\tau}
\int_0^1 \frac{dv}{v} \left\{ \hat{\tau}^4 \frac{{\cal A}^\Phi_{ggg} (S,T,U)}
{\left| \sum_Q F_0^\Phi(\tau_Q) \right|^2}
- 1-\hat\tau^4 - (1-\hat\tau)^4 \right\}
\end{eqnarray}
with $F_0^\Phi$ given in eq.~(A.5). With the functions $f$ and
$g$ given in eq.~(A.7), $A^\Phi_{qqg}$ and $A^\Phi_{ggg}$ can be represented as
\begin{eqnarray}
{\cal A}^H_{qqg} (S) & = & \frac{8}{S-\rho}\left\{ 1-2S~\frac{g(\tau_s)
- g(\tau_Q)}{S-\rho} - \left( 1+\frac{4}{S-\rho}\right)
\left[ f(\tau_s) - f(\tau_Q) \right] \right\} \nonumber \\ \nonumber \\
{\cal A}^A_{qqg} (S) & = & \frac{16}{3(S-\rho)} \left[ f(\tau_s)
- f(\tau_Q) \right] \nonumber \\ \nonumber \\
{\cal A}^{\Phi}_{ggg} (S,T,U) & = & |C_1^{\Phi}|^2 + |C_2^{\Phi}|^2 +
|C_3^{\Phi}|^2 + |C_4^{\Phi}|^2
\end{eqnarray}
The functions $C_i^{\Phi}$ follow from
\begin{equation}
C^\Phi_2(S,T,U) = -C^\Phi_1(T,S,U) \hspace{2cm}
C^\Phi_3(S,T,U) = C^\Phi_1(U,T,S)
\end{equation}
\begin{eqnarray}
C_i^{\Phi} = \sum_Q \frac{1}{2\rho^2}~\sum_{j=1}^{12} P_{ij}^{\Phi} T_j
\end{eqnarray}
The coefficients $T_i$ read
\begin{eqnarray}
T_1 &=& 1 \ \ , \ \ \
T_2 = 2 f(\tau_Q) \ \ , \ \ \
T_3 = 2 f(\tau_s) \ \ , \ \ \
T_4 = 2 f\left(\tau_t \right) \\
T_5 &= & 2 f\left(\tau_u \right) \ \ , \ \ \
T_6 = 2 [1-g(\tau_Q)] \ \ , \ \ \
T_7 = 2 \left[ 1-g(\tau_s)\right] \ \ ,  \ \ \
T_8 = 2 \left[1-g\left(\tau_t \right) \right] \nonumber \\
T_9 & = & 2 \left[1-g\left(\tau_u \right) \right] \ ,   \
T_{10} = J(S,T,U) \ , \
T_{11} = J(S,U,T) \ , \
T_{12} = J(T,S,U) \nonumber
\end{eqnarray}
with
\begin{eqnarray}
J(S,T,U) & = & I_3(S,T,U,S) + I_3(S,T,U,U) - I_3(S,T,U,\rho) \nonumber \\
I_3(S,T,U,X) & = & \frac{1}{SU} \frac{2}{\beta_+
-\beta_-} \left\{ Li_2\left(\frac{\beta_-}{\beta_--\alpha_-}\right)
-Li_2\left(\frac{\beta_+}{\beta_+-\alpha_+}\right) \right. \\
&+ & \left. Li_2\left(\frac{\beta_-}{\beta_--\alpha_+}\right)
-Li_2\left(\frac{\beta_+}{\beta_+-\alpha_-}\right)
+ \log\left(-\frac{\beta_+}{\beta_-} \right)
\log\left(1+\frac{XT}{SU} \right) \right\}  \nonumber
\end{eqnarray}
and
\begin{eqnarray}
\alpha_\pm =  \frac{1}{2} \left( 1 \pm \sqrt{1-\frac{4}{X}} \right)
\ \ , \ \ \ \mbox{\rm and} \ \ \
\beta_\pm = \frac{1}{2} \left( 1 \pm \sqrt{1+\frac{4T}{SU}} \right)
\end{eqnarray}
The coefficients $P_{ij}$ for ${\cal CP}$--even neutral Higgs bosons
are given \cite{P27A} by
\begin{eqnarray*}
P_{1,1} & = & -12~S~\frac{UT - S^2}{(U+S)(T+S)} \\ \\
P_{1,2} & = & 3 \left\{ 4U^3T^3 + 8U^3T^2S + 4U^3TS^2 + 8U^2T^3S +
              15U^2T^2S^2 + 4U^2T^2S \right. \\
& &  + 8U^2TS^3 + 8U^2TS^2 + U^2S^4 -
          4U^2S^3 + 4UT^3S^2 + 8UT^2S^3 \\
& & + 8UT^2S^2 + 8UTS^4 +
         16UTS^3 + 4US^5 - 8US^4 + T^2S^4 - 4T^2S^3 \\
& & \left. + 4TS^5 - 8TS^4 + 3S^6 - 12S^5 \right\} \times
 \frac{1}{S(U+S)^2(T+S)^2}~ \\
P_{1,3} & = & -3~(S-4) \\ \\
P_{1,4} & = & -3~\frac{4U^3T + 8U^2TS - U^2S^2 + 4U^2S + 4UTS^2
              + 8US^2 + S^4 -4S^3}{S(U+S)^2} \\ \\
P_{1,5} & = & -3~\frac{4UT^3 + 8UT^2S + 4UTS^2 - T^2S^2 + 4T^2S
              + 8TS^2 + S^4 -4S^3}{S(T+S)^2} \\ \\
P_{1,6} & = & -12~UT~\frac{U^2T + 2U^2S + UT^2 + 4UTS + 5US^2
              + 2T^2S + 5TS^2 + 4S^3}{(U+S)^2(T+S)^2}
\end{eqnarray*}
\vskip-0.7cm
\parbox{7.2cm}{\begin{eqnarray*}
P_{1,7} & = & 0 \hspace{4.2cm} \\
P_{1,9} & = & 12~UT~(T + 2S)(T+S)^{-2} \\
P_{1,11} & = & 3~TS~(4-S)/2 \\
P_{4,1} & = & 12~\rho \\
P_{4,3} & = & 3~(4-\rho) \\
P_{4,5} & = & 3~(4-\rho) \\
P_{4,7} & = & 0 \\
P_{4,9} & = & 0 \\
P_{4,11} & = & 3~TS~(4-\rho)/2 \\
\end{eqnarray*}}
\hfill
\parbox{7.8cm}{\begin{eqnarray*}
P_{1,8} & = & 12~UT~(U + 2S)(U+S)^{-2} \\
P_{1,10} & = & 3~US~(4-S)/2 \\
P_{1,12} & = & -3~UTS^{-1} (4UT - S^2 + 12S)/2 \\
P_{4,2} & = & -9~(4-\rho) \\
P_{4,4} & = & 3~(4-\rho) \\
P_{4,6} & = & 0 \\
P_{4,8} & = & 0 \\
P_{4,10} & = & 3~US~(4-\rho)/2 \\
P_{4,12} & = & 3~UT~(4-\rho)/2 \\
\end{eqnarray*}}
\vskip-0.3cm
\noindent
Similarly for ${\cal CP}$--odd neutral Higgs bosons, \\
\parbox{4.5cm}{\begin{eqnarray*}
P_{1,1} & = & 0 \hspace{2.5cm} \\
P_{1,3} & = & -2S \\
P_{1,6} & = & 0 \\
P_{1,7} & = & 0 \\
P_{1,9} & = & 0 \\
P_{1,11} & = & -TS^2 \\
P_{4,1} & = & 0 \\
P_{4,3} & = & -2\rho \\
P_{4,5} & = & -2\rho \\
P_{4,7} & = & 0 \\
P_{4,9} & = & 0 \\
P_{4,11} & = & -\rho TS \\
\end{eqnarray*}}
\hfill
\parbox{10.5cm}{\begin{eqnarray*}
P_{1,2} & = & -2S(UT - US - TS - 3S^2)(U+S)^{-1}(T+S)^{-1}  \\
P_{1,4} & = & 2S~(U-S)(U+S)^{-1} \\
P_{1,5} & = & 2S(T-S)(T+S)^{-1} \\
P_{1,8} & = & 0 \\
P_{1,10} & = & -US^2 \\
P_{1,12} & = & STU \\
P_{4,2} & = & 6\rho \\
P_{4,4} & = & -2\rho \\
P_{4,6} & = & 0 \\
P_{4,8} & = & 0 \\
P_{4,10} & = & -\rho US \\
P_{4,12} & = & -\rho UT \\
\end{eqnarray*}}

For the QCD corrections to the gluonic decays of the Higgs bosons
the correction factors $\Delta E_\Phi$ defined in eqs.~(23) and
(70) can be written as
\begin{eqnarray}
\Delta E_\Phi & = & \Delta E_{virt}^\Phi + \Delta E_{ggg}^\Phi + N_F
\Delta E_{gq\bar q}^\Phi \\ \nonumber \\
\Delta E_{virt}^{\cal H} & = & c_{\cal H} (\tau_Q) - \frac{11}{2} \\
\nonumber \\
\Delta E_{virt}^A & = & c_A (\tau_Q) - 6 \\ \nonumber \\
\Delta E_{ggg}^\Phi & = & \int_0^1 dx_1 \int_{1-x_1}^1 dx_2 \left\{
\frac{m_\Phi^6}{stu}
\frac{{\cal A}_{ggg}^\Phi(S,T,U)}{\left| \sum_Q F_0^\Phi(\tau_Q)
\right|^2} - \frac{m_\Phi^8 + s^4 + t^4 + u^4}{2stu~m_\Phi^2} \right\} \\
\nonumber \\
\Delta E_{gq\bar q}^\Phi & = & \int_0^1 dx_1 \int_{1-x_1}^1 dx_2
\frac{s^2+u^2}{t~m_\Phi^2} \left\{ \frac{m_\Phi^2}{s} \left| \frac{\sum_Q
{\cal A}_{qqg}^\Phi (T) }{\sum_Q F_0^\Phi(\tau_Q)} \right|^2 - 1 \right\}
\end{eqnarray}
with $c_{\cal H}$ and $c_A$ given in eqs.~(B.2, B.3); the kinematical variables
are defined as
\begin{equation}
s=m_H^2 (1-x_3) \ , \hspace{0.5cm} t=m_H^2 (1-x_2) \ ,
\hspace{0.5cm} u=m_H^2 (1-x_1) \ , \hspace{0.5cm} x_1+x_2+x_3=2
\end{equation}
and
\begin{equation}
S    = \frac{s}{m_Q^2} \ , \hspace{1cm}
T    = \frac{t}{m_Q^2} \ , \hspace{1cm}
U    = \frac{u}{m_Q^2}
\end{equation}
for the Mandelstam variables normalized by the quark mass.

\newpage

\noindent
{\Large \bf Figure Captions}

\begin{itemize}

\item[{\bf Fig.~1:}]
Generic diagrams of the gluon fusion mechanism $gg\to H$ for the
production of Higgs bosons: lowest order amplitude (a), and
QCD radiative corrections (b).

\item[{\bf Fig.~2:}]
Generic diagrams for the amplitude of the Higgs boson decay into two
photons $H\to\gamma\gamma$:
(a) lowest order $W$--boson amplitude,
(b) lowest order quark amplitude,
and (c) QCD radiative corrections to the quark amplitude.

\item[{\bf Fig.~3:}]
Generic Feynman diagrams for the amplitude of the Higgs boson decay into
gluons $H \to gg$: (a) lowest order amplitude, and (b) QCD
radiative corrections.

\item[{\bf Fig.~4:}]
The real and imaginary parts of the lowest order amplitudes  $A_f$ (a) and
$A_W$
(b) of the  $H \gamma \gamma$ vertex as a function of $\tau_{f,W}=m_H^2/4
m_{f,W}^2$.

\item[{\bf Fig.~5:}]
Real and imaginary parts of the radiative correction factor  to the quark
amplitudes for the $H\gamma\gamma$ coupling; the renormalization point for the
quark mass is taken to be $\mu_Q=m_H/2$ in (a) and $\mu_Q=m_Q$ in (b).

\item[{\bf Fig.~6:}]
(a) The QCD corrected partial decay width of the Higgs boson to two photons
as a function of the Higgs mass, and (b) the size of the QCD
radiative correction factor (in \%).

\item[{\bf Fig.~7:}]
(a) Comparison of the size of the infrared regularized virtual QCD corrections
to the quark amplitude for the pole mass $m_Q(m_Q)$ and the running quark mass
$m_Q(m_H/2)$; for large quark--loop masses the coefficient $C$
approaches the value $C = \pi^2/2 + 11/4$;
(b) The deviation $\Delta E$ of the radiative QCD correction to the decay
$H\to gg$ from its value in the heavy quark--loop limit; the renormalization
scale is taken to be $\mu=m_H$.

\item[{\bf Fig.~8:}]
(a) The QCD corrected  partial decay width of the Higgs boson into two gluons
(in \MeV) as a function of the Higgs mass, and (b) the size of the QCD
radiative
correction factor; the renormalization scale is taken to be  $\mu=m_H$.

\item[{\bf Fig.~9:}]
(a) Total decay width (in \GeV) of the Standard Model Higgs boson as a function
of its mass, and (b) the branching ratios (in \%) of the dominant decay modes
($m_t=174~\GeV$).  All known QCD and leading electroweak radiative corrections
are included.

\item[{\bf Fig.~10:}]
(a) Feynman diagram for the effective couplings of the Higgs boson to
gluons in the heavy--quark--loop limit, and (b) generic Feynman diagrams of
the effective QCD corrections to the decay $H\to gg$ in the heavy--quark--loop
limit.

\item[{\bf Fig.~11:}]
$K$ factors of the QCD corrected cross section $\sigma (pp\to H+X)$ at the LHC
with c.m. energy $\sqrt{s}=14$ \TeV.  $K_{virt}$ and $K_{AB}$ ($A,B = q,g$)
are the regularized virtual correction and the real correction factors,
respectively;  $K_{tot}$ is the ratio of the QCD corrected total cross section
to the lowest order cross section.  The renormalization and factorization
scales are taken to be $\mu=M=m_H$ and the GRV parameterizations for the
parton densities have been used.

\item[{\bf Fig.~12:}]
(a) The spread of the Higgs boson production cross section at the LHC with
c.m. energy of $\sqrt{s}=14$ \TeV~for two parameterizations of the parton
densities. (b)
The total Higgs production cross section at the LHC for two different
c.m.~energy values: $\sqrt{s}=14~\TeV$ and $\sqrt{s} = 10~\TeV$.

\item[{\bf Fig.~13:}]
The renormalization and factorization scale dependence of the Higgs
production cross section at lowest and next--to--leading order; the Higgs mass
is chosen to be (a) $m_H=150~\GeV$, and (b) $m_H = 500~\GeV$.

\item[{\bf Fig.~14:}]
(a) The upper limit of the lightest scalar Higgs boson mass in the $\MSSM$ as
a function of the top quark mass for two values of $\tb= 1.5$ and 30; the top
quark and the common squark masses are taken to be $m_t=174$ \GeV~and $M_S=1$
\TeV, respectively. The dashed line corresponds to the case where $A_t=A_b=\mu
=0$ (only the leading radiative correction is included), while the full lines
correspond to the case where $A_t=A_b=1$ \TeV~and $\mu=-200,0,+200$
\GeV~(from top
to bottom). The masses of the $\MSSM$ Higgs bosons $h^0$, $H^0$ and $H^\pm$,
as a function of the pseudoscalar Higgs mass for the two previous
values of $\tb$ and for $A_t=A_b=\mu=0, M_S=1$ \TeV, are displayed in (b), (c)
and (d), respectively.

\item[{\bf Fig.~15:}]
The coupling parameters of the \MSSM~neutral Higgs bosons as functions of the
pseudoscalar $A^0$ Higgs mass for two values of $\tb=1.5$  and 30 and for
$A_t=A_b=\mu=0, M_S=1$ \TeV. The couplings are
normalized to the \SM~couplings as defined in Table \ref{tb:hcoup}.

\item[{\bf Fig.~16:}]
The branching ratios of the \MSSM~Higgs bosons $h^0$ (a),  $H^0$ (b), $A^0$ (c)
and $H^\pm$ (d) as functions of their masses for two  values of $\tb=1.5$ and
30; the values $A_t=A_b=\mu=0$ and $M_S=1$ \TeV~have been chosen. [The arrows
in (a) denote the branching ratios in the \SM~limit of large $A^0$ masses.]

\item[{\bf Fig.~17:}]
The total decay widths of the \MSSM~Higgs bosons $h^0$,  $H^0$, $A^0$ and
$H^\pm$ as functions of their masses for two values of  $\tb=1.5$ (a) and
$\tb=30$ (b); the values $A_t=A_b=\mu=0$ and $M_S=1$ \TeV~have been chosen.

\item[{\bf Fig.~18:}]
Real and imaginary parts of the QCD radiative correction factor to the
quark amplitudes of the two--photon couplings for
the $\MSSM$ neutral Higgs bosons: (a) $h^0$ and $H^0$ and (b) $A^0$;
the renormalization scale for the quark mass is taken to be $\mu_Q=m_\Phi/2$.

\item[{\bf Fig.~19:}]
The QCD corrected partial decay widths into two photons of the
\MSSM~Higgs  bosons $h^0,H^0,A^0$ for (a) $\tb =1.5$ and (b)
$\tb=30$, and the the size of the QCD
radiative  corrections to the processes $h^0/H^0 \to \gamma\gamma$  and
$A^0  \to\gamma\gamma$ (in \%) as functions of the Higgs boson masses for two
values of $\tb=1.5$ (c) and 30 (d). The renormalization scale for the quark
mass is taken to be $\mu_Q=m_\Phi/2$.

\item[{\bf Fig.~20:}]
The deviation $\Delta E_{\cal H}$ (a) and $\Delta E_A$ (b) of the coefficients
$E_\Phi$ of the radiative QCD correction factors to the process $\Phi\to gg$
from their values in the heavy quark limit, for two values of $\tb=1.5$ and 30;
the renormalization scale is taken to be $\mu=m_\Phi$.

\item[{\bf Fig.~21:}]
The QCD corrected gluonic partial decay widths of the \MSSM~neutral Higgs
bosons $h^0,H^0$ (a) and $A^0$ (b), for two values of $\tb=1.5$ and 30;
the size of the QCD radiative correction factor for $h^0/H^0 \to gg$ (c) and
$A^0\to gg$ (d). The renormalization scale is
taken to be $\mu=m_\Phi$.

\item[{\bf Fig.~22:}]
$K$ factors of the QCD corrected cross sections  $\sigma (pp\to h^0/H^0 +X)$
(a)
and $\sigma(pp\to A^0+X)$ (b) for two values of $\tb=1.5$ and $30$; $K_{virt}$
and $K_{AB}$ ($A,B = q,g$) are the  regularized virtual correction and real
correction factors, respectively, and $K_{tot}$ is the ratio of the QCD
corrected total cross section to the  lowest order cross section. The
renormalization and factorization scales are taken to be  $\mu=M=m_\Phi$ and
the GRV parameterization for the parton densities have been used.

\item[{\bf Fig.~23:}]
The dependence of the total $K$ factors for the processes $\sigma(pp\to
\Phi+X)$ on the value of $\tb$  for a characteristic set of Higgs boson masses.

\item[{\bf Fig.~24:}]
The spread of the \MSSM~Higgs production cross sections $\sigma(pp\to h^0/H^0
+ X)$ (a) and $\sigma(pp\to A^0+X)$ (b) for two parameterizations of the parton
densities.

\item[{\bf Fig.~25:}]
The total production cross sections of the scalar ${\cal  CP}$--even Higgs
bosons $h^0,H^0$ (a) and the pseudoscalar Higgs boson $A^0$ (b) at the LHC
for two
different c.m.~energy values: $\sqrt{s}=14~\TeV$ and $\sqrt{s} =  10~\TeV$.

\item[{\bf Fig.~26:}]
The renormalization/factorization scale dependence of the \MSSM~Higgs boson
production cross sections at lowest and next--to--leading order, for a
characteristic set of Higgs boson masses and $\tb$ values.

\end{itemize}
\end{document}